\documentclass[a4paper,11pt,oneside]{book}
\usepackage{multicol}
\usepackage{float}
\usepackage{graphicx}
\usepackage{lscape}
\usepackage{tikz}
\usepackage{amsmath}
\usepackage{mathtools}
\usepackage{multirow}
\usepackage{hyperref}
\usepackage{authblk}

\title{mcmsupply: An R package for estimating modern contraceptive method supplies} 
\author{Hannah Comiskey and Niamh Cahill}
\affil{Hamilton Institute and Department of Mathematics \& Statistics, Maynooth University, Maynooth, Ireland}

\begin{document}

\maketitle

\chapter[mcmsupply]{mcmsupply: Estimating modern contraceptive method supplies}

\section{Abstract}
In this paper, we introduce the R package \textit{mcmsupply} which implements Bayesian hierarchical models for estimating and projecting modern contraceptive method supply shares over time. The package implements four model types. These models vary by the administration level of their outcome estimates (national or subnational estimates) and dataset type utilised in the estimation (multi-country or single-country contraceptive market supply datasets). \textit{mcmsupply} contains a compilation of national and subnational level contraceptive source datasets, generated by IPUMS and Demographic and Health Survey microdata. We describe the functions that implement the models through practical examples. The annual estimates and projections with uncertainty of the contraceptive market supply, produced by mcmcsupply at a national and subnational level, are the first of their kind. These estimates and projections have diverse applications, including acting as an indicator of family planning market stability over time and being utilised in the calculation of estimates of modern contraceptive use. 

\section{Introduction}

Family Planning 2030 (FP2030) is a ‘global movement dedicated to advancing the rights of people everywhere to access reproductive health services safely and on their own terms’ \cite{FP2030&UnitedNationsFoundation2021}. One step towards achieving this goal is to quantify \textit{how} people are accessing their modern contraceptive supplies.  To date, obtaining estimates of modern contraceptive supply shares in low- and middle-income countries has relied on large-scale national surveys like the Demographic and Health Surveys (DHS). However, these DHS are not annually available and in practice, most countries carry out DHS every 3 to 5 years approximately, with some countries having fewer surveys than this \cite{TheDHSSurveyTypes}. In previous work, we described a model that provides probabilistic estimates of the contraceptive supply share over time with uncertainty and examined the model performance at the national administration division for countries that are participating in FP2030 and have varying amounts of DHS data available \cite{Comiskey2022}. The original modern contraceptive supply share model (mcmsupply model) relies on splines, informed by cross-method correlations, to capture temporal variation combined with a hierarchical modelling approach to estimate country-level parameters. Using a multi-country dataset, the original mcmsupply model produces estimates of contraceptive supply market shares at the national level for all countries simultaneously. In this paper, we extend the model to estimate supply shares using input data from a single-country and to also include estimation at the subnational administrative division. 

\begin{figure}[H]
    \centering
    \includegraphics[width=15cm]{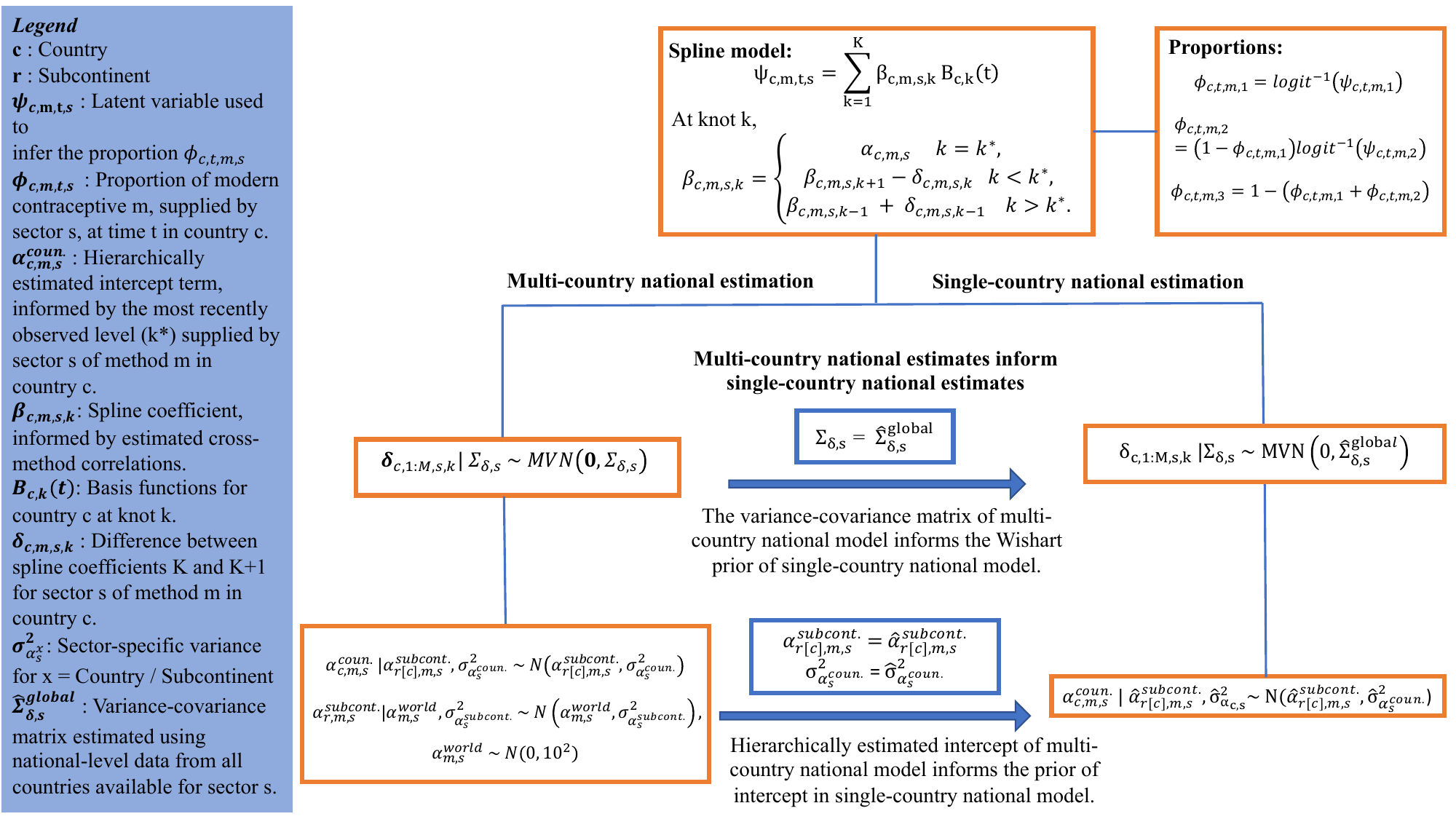}
    \caption{A flow chart to illustrate the relationship between the multi-country and single-country national estimation approaches. The median estimates of the multi-country national model parameters are used as informative priors in the single-country national model. A legend describing each element of the schematic is located in the blue box.}
    \label{fig:fig_3_1}
\end{figure}

For the remainder of the paper, we will refer to the original modern contraceptive supply share model as the multi-country national mcmsupply model. The single-country \textit{mcmsupply} model uses a scaled-down version of the multi-country approach. It borrows strength from the multi-country model using modular model runs with informative priors placed on key parameters to provide precise outcome estimation, even in the absence of data for a particular contraceptive method. Modularization in Bayesian analysis describes the process within a statistical model where information is restricted to flow only from the prior to the likelihood. Thus, preventing the ‘contamination’ of key parameters from suspect data \cite{Plummer2015CutsModels} \cite{Lunn2009}. In the context of our problem, parameters estimated within the multi-country model are used to inform the priors of the single-country (either national or subnational administrative division) models. This approach prevents spurious parameter estimates due to a lack of data for some counties.  To summarise how the single-country and multi-country models are connected to each other, Figure \ref{fig:fig_3_1} depicts this modelling relationship at the national administration level. The main differences between the multi-country and single-country approaches is that for a single-country model, we only have data for one country (at the national or subnational administration division) and the country-level (in the national model) or subnational-level (in the subnational model) population parameters are informed by estimates from the corresponding multi-country model. In contrast to this, the multi-country model uses national or subnational level data (depending on your administrative level of interest) from many countries simultaneously to estimate model parameters and the country-level (in the national model) or subnational-level (in the subnational model) population parameters are estimated hierarchically.

The need for the single-country version of the mcmsupply model arose for two primary reasons. Firstly, there was a demand for improved computational efficiency while ensuring that model accuracy remains uncompromised when generating projections of modern contraceptive method supply shares for a single-country. Secondly, the option to accommodate custom data, like incorporating a new survey dataset, was sought to be included in the estimation process for users who require a more tailored analysis.

The inclusion of the functionality to estimate supply shares at the subnational level in the mcmsupply package was spurred by the growing interest in subnational estimation among the family planning community \cite{New2017}, \cite{Mercer2019}, \cite{Li2019}. The decentralization of family planning services produces a more equitable and efficient service; however, it also shifts the responsibility of service delivery to lower-level organisations that may not have the capacity to carry-out the role \cite{Williamson2014}. Providing subnational-level estimates can lead to a clearer understanding of localised user preferences, localised user access to family planning commodities and a measure of the true stability contraceptive supply market at a smaller geographic scale \cite{BOSSERT2002}.  These subnational estimates may also be used as part of a localised temperature check for progress towards the FP2030 goals, which include increasing access to contraceptive methods \cite{Munoz2022}. 

\begin{figure}[H]
    \centering
    \includegraphics[width=15cm]{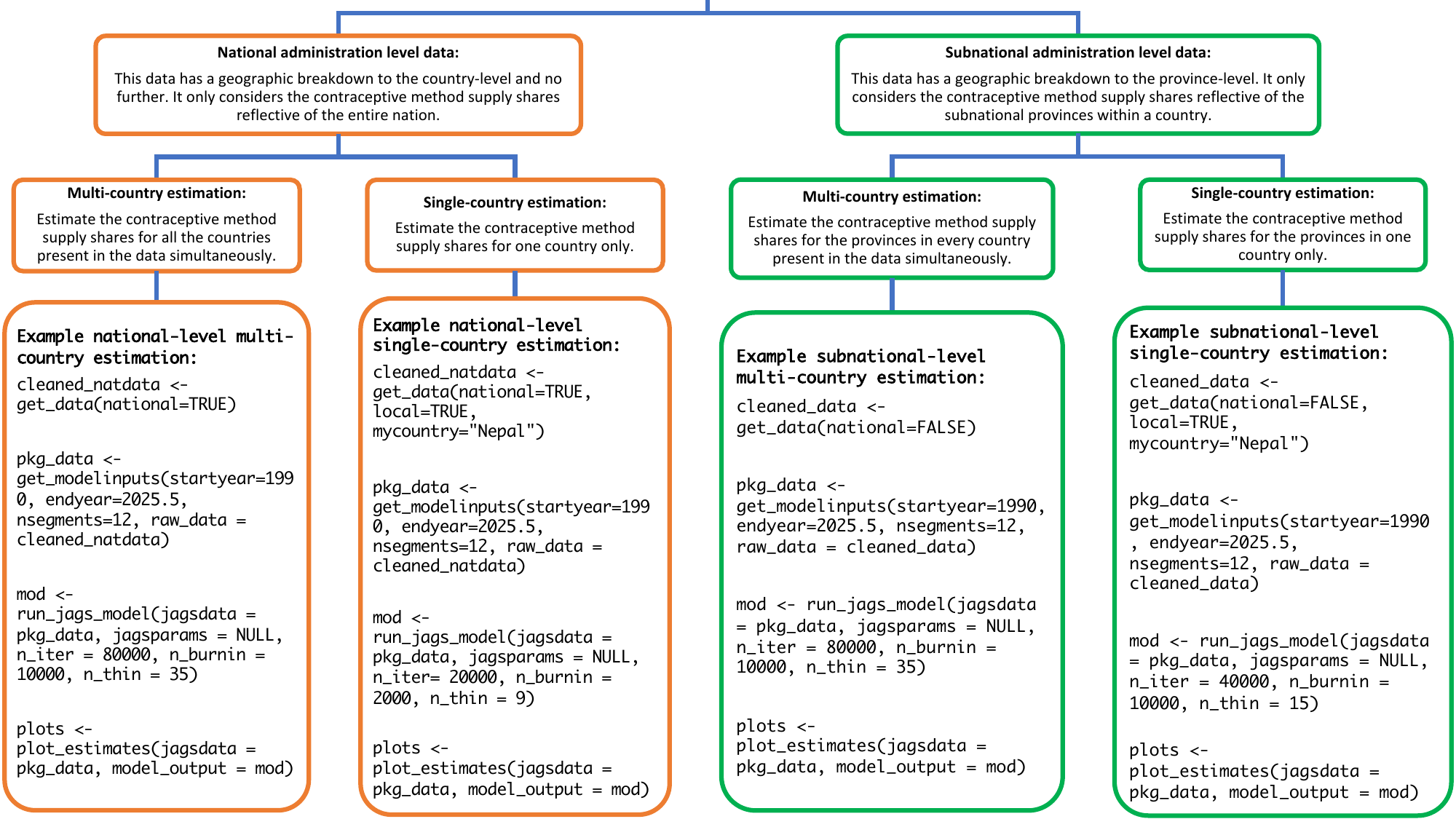}
    \caption{A flow chart to illustrate the decision processes that lead to the different estimation types within the mcmsupply package. The first decision is with respect to the administrative level of the estimates you wish to create – they may be either national level or subnational level. An explanation of each division is found on the first row of the figure. The second decision is with respect to the number of countries you wish to estimate – the user may estimate the proportions for all the countries at once or only one country. An explanation of each in the context of the specific administrative division is found on the second row of the figure. A set of sample functions used to estimate and plot the estimates for each modelling option are located on the third row of the figure}
    \label{fig:fig_3_2}
\end{figure}

This paper introduces the R package \textit{mcmsupply} for estimating and projecting the contraceptive supply share at the national and subnational administration divisions using multi-country or single-country datasets. Figure \ref{fig:fig_3_2} shows a summary of the different ways users can use the R functions and input data in \textit{mcmsupply} to do model fitting and estimation. The national data contained in the package is derived from the DHS microdata \cite{ICF2012}  while the subnational data is derived from the IPUMS DHS datasets \cite{IPUMSdata}. In the 'Implementation and Operation' section, we review the operation and implementation requirements of the \textit{mcmsupply} R package. 

In the 'Data' section, a detailed description of the data used within the \textit{mcmsupply} R package is provided, as well as an explanation of the data pre-processing functions \verb|get_data| and \verb|get_modelinputs|. 'The estimation process' section describes the model fitting and visualisation functions within the \textit{mcmsupply} R package. A basic overview of the process model is provided in 'Overview of the mcmsupply process models' with an explanation of some key modelling parameters. The 'Model fitting' section  explains the \verb|run_jagsmodel| function. The  \verb|run_jagsmodel| function fits models in a Bayesian framework using the JAGS (Just Another Gibbs Sampler) software and produces estimates with uncertainty for the administration level and dataset type of choice \cite{Plummer2003}. The 'Model output' section describes the \verb|plot_estimates| function that takes the model estimates and visualises them using the R package ggplot2 \cite{Wickham2016}. The 'Use cases' section discusses five use cases for modelling modern contraceptive supply shares using the \textit{mcmsupply} R package. Finally, the conclusions are presented.

\section{Implementation and Operation}

\textit{mcmsupply} contains pre-processing functions to clean and prepare raw input data for model fitting at the administrative level of choice (national or subnational) using the dataset type of choice (multi-country or single-country).  This R package includes functions to fit Bayesian hierarchical models using the model inputs. Functionality for post-processing and visualisation of the model estimates are also included. The model fitting process described uses JAGS. JAGS uses Markov Chain Monte Carlo (MCMC) sampling to produce model estimates for Bayesian hierarchical models \cite{Plummer2003}. For installation, both R (>=3.5.0) and JAGS (>=4.0.0) are required. JAGS can be downloaded at https://sourceforge.net/projects/mcmc-JAGS/files/JAGS/. \textit{mcmsupply} interacts with JAGS using the wrapper functions supplied by the R package \textit{R2jags} \cite{Su2021}. The \textit{mcmsupply} package dependencies are listed in the package \verb|DESCRIPTION| file and will be automatically installed upon installing the main package. There are no minimum RAM, CPU, or HARDDRIVE requirements apart from what is necessary to store model runs, which varies case-by-case. This software was run on a MacBook Air using macOS 13.1 with a 1.6 GHz Dual-Core Intel Core i5 processor and 8GB of memory.

\section{Data}

\subsection{Input data}

The mcmsupply package contains the following input data sources: 
\begin{enumerate}
\item Survey data for the national and subnational modern contraceptive supply shares between 1990 and 2020 for a selection of countries participating in the FP2030 initiative. 
\item Estimated correlations between the rates of change in method supply shares in the public and private sectors at both the national and subnational administrative divisions. These are derived based on the method outlined in Comiskey et al., 2023. 
\item Global and subcontinental parameter estimates obtained from the multi-country model run at the national and subnational level.
\end{enumerate}

The inst/data-raw folder of the \textit{mcmsupply} package contains sample R code that was used to create the model inputs in the case of the correlations and parameters. This enables users to recalculate their own single-country model parameters and correlations should they wish to do so. The code for the creation of the main input contraceptive supply source data is provided on the \textit{mcmsupply} github page but the raw data for these datasets cannot be provided. The raw data may be accessed by users through an application to the DHS program for national level data, or the IPUMS program, for subnational level data. Help files for the contraceptive supply source datasets can be accessed by using the command \verb|?mcmsupply::national_FPsource_data| and \verb|?mcmsupply::subnat_FPsource_data|. The national contraceptive supply source data has data for 67 countries (including participants and non-participants of FP2030) between 1990 and 2020. The subnational contraceptive supply source data has data for 246 provinces across 24 countries (all participating in FP2030) between 1990 and 2020. Table \ref{tab:tab3_1} is a sample of 6 rows of the subnational contraceptive supply source data.

\begin{table}[H]
\resizebox{\textwidth}{!}{%
\begin{tabular}{|c|c|c|c|c|c|c|c|}
\hline
\textbf{Country} & \textbf{Region} & \textbf{Method} & \textbf{average\_year} & \textbf{sector\_categories} & \textbf{proportion} & \textbf{SE.proportion} & \textbf{n} \\ \hline
Zimbabwe & Midlands & OC Pills & 2010.5 & Commercial\_medical & 0.16017376 & 0.03557217 & 40  \\ \hline
Zimbabwe & Midlands & OC Pills & 2010.5 & Other               & 0.10185427 & 0.02596022 & 29  \\ \hline
Zimbabwe & Midlands & OC Pills & 2010.5 & Public              & 0.73797197 & 0.04498512 & 194 \\ \hline
Zimbabwe & Midlands & OC Pills & 2015.5 & Commercial\_medical & 0.19294664 & 0.03954100 & 60  \\ \hline
Zimbabwe & Midlands & OC Pills & 2015.5 & Other               & 0.04282267 & 0.01660390 & 13  \\ \hline
Zimbabwe & Midlands & OC Pills & 2015.5 & Public              & 0.76423069 & 0.03875660 & 209 \\ \hline
\end{tabular}%
}
\caption{The subnational contraceptive supply source data used in the subnational estimation models. Country and Region list the name of the country and province the observation relates to. The Method column lists the type of the contraceptive method supplied. The mid-year when the survey was collected is listed in average\_year. The supply sector is found in sector\_categories. The observed proportion and standard error are found in proportion and SE.proportion. The number of respondent making up each observation are listed in n}
\label{tab:tab3_1}
\end{table}

Lastly, data on country classification, ISO codes and area groupings is provided in the dataset \verb|Country_and_area_classification| (Table \ref{tab:tab3_2}). The help file for this dataset can be accessed via the R command \\ \verb|?mcmsupply::Country_and_area_classification|.

\begin{table}[H]
\resizebox{\textwidth}{!}{%
\begin{tabular}{|c|c|c|c|c|c|c|c|}
\hline
\textbf{Country or area} &
  \textbf{ISO Code} &
  \textbf{Major area} &
  \textbf{Region} &
  \textbf{Developed region} &
  \textbf{Least developed country} &
  \textbf{Sub-Saharan Africa} &
  \textbf{FP2020} \\ \hline
Afghanistan    & 4  & Asia    & Southern Asia   & No  & Yes & No & Yes \\ \hline
Albania        & 8  & Europe  & Southern Europe & Yes & No  & No & No  \\ \hline
Algeria        & 12 & Africa  & Northern Africa & No  & No  & No & No  \\ \hline
American Samoa & 16 & Oceania & Polynesia       & No  & No  & No & No  \\ \hline
\end{tabular}%
}
\caption{The Country\_and\_area\_classification dataset is the Track20 project country and area classification data according to the United Nations Statistical Division, standard country or area codes for statistical use (M49). This data set is how we classify each country in subcontinental regions. The name of the country, the International Organization for Standardization (ISO) code for each country, the continent, sub-continent are listed. Details on whether or not a country is defined as a developing, located in Sub-Saharan Africa and the status of it's participation in FP2020 (now FP2030) are also provided.}
\label{tab:tab3_2}
\end{table}

\subsection{Data pre-processing}

In \textit{mcmsupply}, the data processing occurs in two steps: First, the raw input data is retrieved and preliminary cleaning to the dataset is completed. Secondly, the cleaned data is processed to provide the model inputs for the Bayesian hierarchical model. This two-step process removes any black-box element to the model fitting process and allows the user to review the data at both stages and refer to it later when considering model outputs.
The first step involves the \verb|get_data| function. This function retrieves the raw data from the stored contraceptive supply source dataset, does data cleaning and processing to address any issues with missing data and regional naming inconsistencies. Its arguments are summarized in Table \ref{tab:tab3_3}. First, the user defines whether they wish to use national or subnational level administrative data via the \verb|national| argument. National level data is accessed when the \verb|national| argument is set to \verb|TRUE|. When subnational administrative data is required, the user sets national to \verb|FALSE|. Similarly, the user defines whether they want to use multi-country estimation with data from multiple countries or single-country estimation with data from a single-country via the \verb|local| argument. The default setting for the \verb|local| argument is \verb|FALSE|. This induces a multi-country estimation, where the outcomes for all the countries in the contraceptive supply source dataset will be estimated simultaneously. In the event of single-country estimation, the user sets \verb|local| to \verb|TRUE| and indicates their country of interest via the \verb|mycountry| argument. The names of the countries listed in the package data can be found in the \verb|country_names| dataset. The help file for this dataset can be accessed via the R command \verb|?mcmsupply::country_names|. The \verb|fp2030| argument controls whether to include countries that are participating in the FP2030 initiative or not. The default includes only the named FP2030 countries (see \verb|country_names|) in the dataset. There is the optional functionality to include a custom dataset. This allows the user to run the model on data outside of that stored within the package. The \verb|surveydata_filepath| is a character string that denotes the location of the custom dataset. The file must meet a series of internal checks on file type, column names, suitable data ranges and missing data. When a custom dataset is supplied to \verb|get_data|, the function carries out the checks and alerts the user to any differences between what is expected and what has been supplied. If \verb|surveydata_filepath| is left as \verb|NULL|, by default the function uses the stored \verb|national_FPsource_data| or \verb|subnat_FPsource_data|, depending on what administrative level the user has specified via the \verb|national| argument (Table \ref{tab:tab3_3}). The \verb|get_data| function returns a list containing the cleaned data and a list of arguments supplied to the function. Storing the arguments of the \verb|get_data| function allows the set-up information to flow without requiring the user to repeatedly supply the same arguments for each step of the modelling process.

\begin{table}[H]
\resizebox{\textwidth}{!}{%
\begin{tabular}{|c|c|c|}
\hline
\textbf{Argument} &
  \textbf{Data type} &
  \textbf{Description} \\ \hline
national &
  Character &
  \begin{tabular}[c]{@{}c@{}}It indicates whether the user is interested in using data at the\\ national or subnational administration level. \\ This is a binary TRUE or FALSE argument. \\ Default is TRUE which retrieves national level data, while \\ FALSE retrieves subnational data.\end{tabular} \\ \hline
local &
  Character &
  \begin{tabular}[c]{@{}c@{}}It indicates whether the user is interested in using data for a\\ single population or not. \\ This is a binary TRUE or FALSE argument. \\ Default is FALSE.\end{tabular} \\ \hline
mycountry &
  Character &
  \begin{tabular}[c]{@{}c@{}}This is the name of the country you wish to do single-country \\ estimation for. The data will only be returned for this country. \\ Default is NULL.\end{tabular} \\ \hline
fp2030 &
  Character &
  \begin{tabular}[c]{@{}c@{}}It indicates whether the user is interested in using only \\ countries participating in FP2030. \\ This is a binary TRUE or FALSE argument. \\ Default is TRUE.\end{tabular} \\ \hline
surveydata\_filepath &
  Character  &
  \begin{tabular}[c]{@{}c@{}}Pathway to the location of the custom dataset. \\ When left as NULL, the function \\ automatically uses the stored datasets.\\ Default is NULL. \end{tabular} \\ \hline
\end{tabular}%
}
\caption{The arguments of the get\_data function. The purpose of this function is to retrieve and clean the Demographic and Health Survey (DHS) data or custom user supplied data for use in supply share estimation. The Argument column names the function component. Data type describes the argument. Description explains the purpose of the argument and any default entries.}
\label{tab:tab3_3}
\end{table}

Step two of the data pre-processing is the \verb|get_modelinputs| function. This function takes the cleaned data from the previous step and repackages it into suitable inputs for the model implementation. The arguments of the function are summarised in Table \ref{tab:tab3_4}. This function uses the arguments set in the \verb|get_data| function as well as additional parameters for the model. These parameters include the year the user wishes to begin their estimation at and the year they finish on.  In the \textit{mcmsupply} package, the models use basis splines (B-splines), to capture the complexities in variation of the contraceptive supply source data over time. B-splines use basis-functions to create piece-wise cubic polynomials. The number of basis functions that are fit to the data is determined by the number of knots. Knots are the locations along the x-axis where the piece-wise polynomials of the B-splines join. As you increase the number of knots in the basis functions, the B-splines give a tighter fit to the data. Similarly, if you decrease the number of knots in the basis, you will get a smoother fit to your data.   In the \textit{mcmsupply} package, the user may alter the number of knots (\verb|nsegments|) used in the basis functions. The default number of knots is 12, as was used in Comiskey et al., 2023.  Like the \verb|get_data| function, this function returns a list containing the model inputs and the function arguments.

\begin{table}[H]
\resizebox{\textwidth}{!}{%
\begin{tabular}{|c|c|c|}
\hline
\textbf{Argument} & \textbf{Data type} & \textbf{Description}                            \\ \hline
startyear         & Numeric            & The year you wish to start your estimation at.  \\ \hline
endyear           & Numeric            & The year you wish to finish your estimation at. \\ \hline
nsegments &
  Numeric &
  \begin{tabular}[c]{@{}c@{}}The number of knots   you wish to include in your basis functions. \\ Default is 12.\end{tabular} \\ \hline
raw\_data &
  List &
  \begin{tabular}[c]{@{}c@{}}The output of the get\_data function, which includes a list \\ of the function arguments used and the \\ cleaned contraceptive supply source data.\end{tabular} \\ \hline
\end{tabular}%
}
\caption{The arguments of the get\_modelinputs function. The purpose of this function is to get the model inputs for the JAGS model used for supply share estimation. In the table, the argument name and data type of the argument is stated, a description of the argument and any default values is then provided.}
\label{tab:tab3_4}
\end{table}

\section{The estimation process}

The \textit{mcmsupply} R package contains four Bayesian models, each of which aims to estimate and project contraceptive method supply shares over time with uncertainty. These models vary by the administration level of their outcome estimates (national or subnational estimates) and dataset type utilised in the estimation (multi-country or single-country contraceptive market supply datasets). A full mathematical description of all the models contained within \textit{mcmsupply} are described in the appendix \footnote{At the time of publication it is expected that this material will be made available via another open access venue.}. A summary of each of the parameters and their role within each model can be found in Table \ref{tab:tab3_5} while a visual summary of the national model, using both multi-country and single-country inputs, can be found in Figure \ref{fig:fig_3_1}.

\subsubsection{Brief model overview}
The outcome of interest is the components of a compositional vector $\boldsymbol{{\phi_{q,t,m}}}$, which captures the proportion of contraceptive method m, at time t, in population q supplied across the public and private sectors.

\begin{equation}
\boldsymbol{{\phi_{q,t,m}}} =(\phi_{q,t,m,1},\phi_{q,t,m,2},\phi_{q,t,m,3}),
\end{equation} 

where, \newline
$\phi_{q,t,m,s}$ is the proportion supplied by the public sector (s =1), the private commercial medical sector (s=2) and the other private sector (s=3) of modern contraceptive method m, at time t, in population q (national or subnational). \newline 

Figure \ref{tab:tab3_1} shows the model set up for the national level models. A similar approach is taken when estimating modern contraceptive method supply at the subnational administration level. For each model within the \textit{mcmsupply} package, the latent variable $\psi_{q,m,t,s}$ relies on a spline to capture the underlying process that generates the data, on the logit scale, for sector s, in year t, for method m and population q (depending on the administration level of interest) .

\begin{equation}
   \psi_{q,t,m,1}=\sum_{k=1}^K \beta_{q,m,1,k} B_{q,k}(t)
\end{equation}

where, \newline
$\beta_{q,m,1,k}$ is the $k^{th}$ spline coefficient for sector s, method m in population q. \newline
$B_{q,k}(t)$ is the $k^{th}$ basis function fit to the data for population q. \newline

We assume that in population q, for method m and sector s, the value of spline coefficient at knot index $k^*$, aligning with the year $t^*$, the most recent survey available, is $\alpha_{q, m, s}$. By doing this, we are assuming that the $\alpha_{q,m,s}$ parameter will act as the spline coefficient for the reference spline at $k^*$. We are then able to calculate the remaining spline coefficients from the reference index ($k^*$) using the estimated $\boldsymbol{\delta_{q,m,s}}$.

\begin{equation}
    \beta_{q,m, s, k}=\left\{\begin{array}{c}
    \alpha_{q,m,s} \quad k=k^*, \\
    \beta_{q,m,s, k+1}-\delta_{q, m, s, k} \quad k<k^*, \\
    \beta_{q,m,s, k-1}+\delta_{q, m, s, k-1} \quad k>k^*,
\end{array}\right.
\end{equation} 

where, \newline
$\alpha_{q,m,s}$ is the most recently observed supply share, on the logit scale, for sector s , method m, in population q. This parameter is estimated hierarchically. The geographical set-up of this estimation process adapts to match the administrative level of interest. For example, the subnational multi-country models contain an additional layer of geography ($world > subcontinent > country > province$) in the hierarchical set-up that the national models don't have, which accounts for the subnational administration levels. \newline
k is the knot index along the set of basis splines $B_{q,k}(t)$ \newline
$k^*$ is the index of the knot that corresponds with $t^*$, the year index where the most recent survey occurred in population q. \newline
$\delta_{q,m,s,k-1}$ is the first order difference between spline coefficients $\beta_{q,m, s, K}$ and $\beta_{q,m, s,K-1}$. These reflect the changes in method supply shares over time. Within each sector, the first-order differences are assumed to be correlated  between methods. These correlations were estimated using a maximum \textit{a posteriori} estimator for the correlation matrix first described in Azose and Raftery, 2018 \cite{Azose2018} and adapted for method supply shares in Comiskey et al.,2023. These estimated correlations are available as data for both the national and subnational models. Please see the Data section of this paper for more details.

\begin{table}[H]
\resizebox{\textwidth}{!}{%
\begin{tabular}{|c|c|c|}
\hline
\textbf{Model type} &
  \textbf{Parameter name} &
  \textbf{Parameter purpose} \\ \hline
\multirow{4}{*}{Multi-country national} &
  P &
  \begin{tabular}[c]{@{}c@{}}The method supply share proportions ($\phi_{c,t,m,s}$) \\ for all countries, methods and sectors.\end{tabular} \\ \cline{2-3} 
 &
  beta.k &
  \begin{tabular}[c]{@{}c@{}}The set of spline coefficients for ($\beta_{c,m,s,k}$) \\ for all countries, methods and sectors at each knot.\end{tabular} \\ \cline{2-3} 
 &
  alpha\_cms &
  \begin{tabular}[c]{@{}c@{}}The intercept term ($\alpha_{c,m,s}$) \\ for all countries, methods and sectors.\end{tabular} \\ \cline{2-3} 
 &
  delta.k &
  \begin{tabular}[c]{@{}c@{}}The   first order differences between spline coefficients (($\delta_{c,m,s,k}$)) \\ across all knots for all countries, methods and sectors\end{tabular} \\ \hline
\multirow{4}{*}{Single-country national} &
  P &
  \begin{tabular}[c]{@{}c@{}}The method supply share proportions ($\phi_{c,t,m,s}$) \\ for all countries, methods and sectors.\end{tabular} \\ \cline{2-3} 
 &
  alpha\_cms &
  \begin{tabular}[c]{@{}c@{}}The intercept term ($\alpha_{c,m,s}$) for \\ the country of interest c, across all methods and sectors.\end{tabular} \\ \cline{2-3} 
 &
  inv.sigma\_delta &
  \begin{tabular}[c]{@{}c@{}}The precision matrix used in the \\ multivariate normal prior of $\delta_{c,1:M,s,k}$\end{tabular} \\ \cline{2-3} 
 &
  beta.k &
  \begin{tabular}[c]{@{}c@{}}The set of spline coefficients ($\beta_{c,m,s,k}$) for \\ the country of interest c, across all methods and sectors at each knot.\end{tabular} \\ \hline
\multirow{6}{*}{Multi-country subnational} &
  alpha\_pms &
  \begin{tabular}[c]{@{}c@{}}The intercept term ($\alpha_{p,m,s}$)  \\ for all subnational provinces, methods and sectors.\end{tabular} \\ \cline{2-3} 
 &
  alpha\_cms &
  \begin{tabular}[c]{@{}c@{}}The intercept term   ($\alpha_{c,m,s}$) \\ for all countries, methods and sectors.   \\ $\alpha_{c[p],m,s}$ is the expected value of $\alpha_{p,m,s}$.\end{tabular} \\ \cline{2-3} 
 &
  inv.sigma\_delta &
  \begin{tabular}[c]{@{}c@{}}The precision matrix used in the \\ multivariate normal prior of  $\delta_{p,1:M,s,k}$\end{tabular} \\ \cline{2-3} 
 &
  tau\_alpha\_pms &
  \begin{tabular}[c]{@{}c@{}}The sector-specific precision\\ associated with $\alpha_{p,m,s}$\end{tabular} \\ \cline{2-3} 
 &
  beta.k &
  \begin{tabular}[c]{@{}c@{}}The set of spline coefficients ($\beta_{p,m,s,k}$) \\ across all subnational provinces, methods and sectors at each knot.\end{tabular} \\ \cline{2-3} 
 &
  delta.k &
  \begin{tabular}[c]{@{}c@{}}The first order differences between spline coefficients \\ across all knots ($\delta_{p,m,s,k}$) for all provinces, methods and sectors.\end{tabular} \\ \hline
\multirow{3}{*}{Single-country subnational} &
  P &
  \begin{tabular}[c]{@{}c@{}}The method supply share proportions ($\phi_{p,t,m,s}$) for the \\ subnational provinces in the country of interest, for all methods and sectors.\end{tabular} \\ \cline{2-3} 
 &
  alpha\_pms &
  \begin{tabular}[c]{@{}c@{}}The intercept term ($\alpha_{p,m,s}$) for the \\ subnational provinces in the country of interest, for all methods and sectors.\end{tabular} \\ \cline{2-3} 
 &
  beta.k &
  \begin{tabular}[c]{@{}c@{}}The set of spline coefficients ($\beta_{p,m,s,k}$) \\ for the subnational provinces in the country of interest, \\ across all methods and sectors at each knot\end{tabular} \\ \hline
\end{tabular}%
}
\caption{This table summarises the purpose of each parameter within each of the four models described in the \textit{mcmsupply} R package. The four types of model are listed in the 'Model type' column. For each model, the parameters listed in the 'Parameter name' column are the default parameters monitored when the argument jagsparams=NULL is used within the run\_jags\_model function. The 'Parameter purpose' column explains the role each parameter plays within the estimation process and the notation can be linked directly to the 'Model overview' section of this paper. }
\label{tab:tab3_5}
\end{table}

\subsection{Model fitting}

The \verb|run_jags_model| function fits the selected JAGS model to the supplied data and returns a list of MCMC samples and point summaries for the time-period and locations of interest. Initial set-up arguments (\verb|national|, \verb|local|, \verb|mycountry|) are inherited from the specification of the previous \verb|get_modelinputs| function. The additional inputs of this function are summarised in Table \ref{tab:tab3_6}. The \verb|jagsdata| argument of this function is a list of initial set-up arguments and JAGS model inputs gathered from the \verb|get_modelinputs| function. \verb|jagsparams| is a vector of strings that name the model parameters the user wishes to monitor within the JAGS model. The default is \verb|NULL|. When \verb|jagsparams = NULL|, the function will refer to a stored vector of parameters to monitor (Table \ref{tab:tab3_5}). The JAGS parameters of \verb|n_iter = 80000|, \verb|n_burnin = 10000| and \verb|n_thin = 35| ensure that the model has converged and that the final posterior sample size is 2000 samples. The function \verb|get_point_estimates| takes the chains produced by the JAGS model and estimates the median, 95\% credible intervals and 80\% credible intervals. The \verb|get_point_estimates| function runs automatically inside the \verb|run_jags_model| function and returns the point summaries as part of the \verb|run_jags_model| function output. The \verb|run_jags_model| function returns a list containing the JAGS output of the model and the point summaries for the estimates.

\begin{table}[H]
\resizebox{\textwidth}{!}{%
\begin{tabular}{|c|c|c|}
\hline
\textbf{Argument} & \textbf{Data type} & \textbf{Description}                                                                                                            \\ \hline
jagsdata &
  List &
  \begin{tabular}[c]{@{}c@{}}The output of the get\_modelinputs function. \\ A list of the initial set-up arguments and the JAGS inputs \\ required using the data.\end{tabular} \\ \hline
jagsparams &
  Vector &
  \begin{tabular}[c]{@{}c@{}}A string vector of model parameters to be monitored. \\ Default is NULL. \\ NULL invokes a standard  vector to be used.\end{tabular} \\ \hline
n\_iter           & Numeric            & \begin{tabular}[c]{@{}c@{}}Number of iterations you wish to run your JAGS model for. \\ Default is 80000.\end{tabular}       \\ \hline
n\_burnin         & Numeric            & \begin{tabular}[c]{@{}c@{}}Number of burn-in   samples you wish to run your JAGS model for. \\ Default is 10000\end{tabular} \\ \hline
n\_thin           & Numeric            & \begin{tabular}[c]{@{}c@{}}Number of samples you wish to thin your JAGS sample by. \\ Default is 35.\end{tabular}            \\ \hline
\end{tabular}%
}
\caption{The arguments of the run\_jags\_model function. The purpose of this function is to run the Bayesian hierarchical models stored within the mcmsupply package for either single- or multi-country datasets at the administration level of interest. The argument name and data type of the argument is stated, a description and any default values of the argument are then provided.}
\label{tab:tab3_6}
\end{table}

\subsubsection{Model output}

The user then runs the function \verb|plot_estimates|. This function visualises the point estimates with uncertainty alongside the data using the initial set up inputs of the \verb|get_modelinputs| function and the output of the \verb|run_jags_model| function (Table \ref{tab:tab3_6}). The \verb|plot_estimates| function returns a list of \textit{ggplot2} objects, one for each country (when using the national model) or subnational region (when using the subnational model).

\begin{table}[H]
\resizebox{\textwidth}{!}{%
\begin{tabular}{|c|c|c|}
\hline
\textbf{Argument} &
  \textbf{Data type} &
  \textbf{Description} \\ \hline
jagsdata &
  List &
  \begin{tabular}[c]{@{}c@{}}A list of the initial set-up arguments and the JAGS inputs required \\ using the data retrieved in the get\_modelinputs function.\end{tabular} \\ \hline
model\_output &
  List &
  \begin{tabular}[c]{@{}c@{}}The object assigned to store the list of  MCMC results and estimate \\ summary output from the run\_jags\_model function.\end{tabular} \\ \hline
\end{tabular}%
}
\caption{The arguments of the plot\_estimates function. The purpose of this function is to plot the data alongside the model estimates so that users can visualise their estimated method supply shares. The argument name and data type of the argument is stated, a description of the argument is then provided. }
\label{tab:tab3_7}
\end{table}

\section{Use cases}

\subsection{Case 1: Estimating contraceptive method supply shares at the national administration level for multiple countries simultaneously}

The first use case describes how the user can estimate modern contraceptive method supply shares at the national administrative level over time for multiple countries at once (i.e., using the multi-country national model). 

This use case is described in \verb|multi_national_mod| found in the \verb|vignettes| folder. This vignette takes approximately 12 hours to run on a machine with 1.6 GHz Dual-Core Intel Core i5 processor and 8GB of RAM. The \verb|national_FPsource_data| dataset contains observations for 30 countries. The user begins by accessing the \verb|national_FPsource_data| dataset through the \verb|get_data| function with the argument \verb|national=TRUE|, and the remaining arguments sets to their default values, to indicate that they are interested in national-level data for the FP2030 countries present in the data.

\begin{verbatim}
cleaned_natdata <- get_data(national=TRUE)
\end{verbatim}

Next, this data is supplied to the \verb|get_modelinputs| function. This function reshapes the data into a list of inputs for the JAGS model. At this point, the user must indicate the start and end years they wish to estimate between. The \verb|n_segments| argument controls how many knots will be used in the basis functions. The default number of segments is 12. Lastly, the cleaned national data from the \verb|get_data| function is provided to the \verb|get_modelinputs| function via the \verb|raw_data| argument.

\begin{verbatim}
pkg_data <- get_modelinputs(startyear=1990, 
                            endyear=2025.5,  
                            nsegments=12,  
                            raw_data = cleaned_natdata)
\end{verbatim}

This list of data and model inputs is then fed into the JAGS model via the \verb|run_jags_model| function. In this instance, the user wishes to monitor the default set of parameters within the JAGS model. Therefore, they set the `jagsparams` argument to `NULL`, which invokes the function to use the default list. The parameters for running the JAGS model are set via the \verb|n_iter|, \verb|n_burnin| and \verb|n_thin| arguments. As part of the \verb|run_jags_model| function, the median and 80\% and 95\% credible intervals for the estimates are calculated.

\begin{verbatim}
mod <- run_jags_model(jagsdata = pkg_data, 
                      jagsparams = NULL,
                      n_iter = 80000, 
                      n_burnin = 10000, 
                      n_thin = 35)
\end{verbatim}

The final JAGS model output and the summary estimates are returned as a list. Finally, these summary estimates are visualised via the \verb|plot_estimates| function using the R package \textit{ggplot2} (Figure \ref{fig:fig3_3}).

\begin{verbatim}
plots <- plot_estimates(jagsdata = pkg_data, 
                        model_output = mod) 
\end{verbatim}

\begin{figure}[H]
    \centering
    \includegraphics[width=15cm]{_figs/fig_3.pdf}
    \caption{The plotted posterior point estimates for each of the three sectors (public in blue, private commercial medical in grey, and private other in gold) for Nepal at the national administrative level over time with the 80\% and 95\% uncertainty interval denoted as shaded regions. The survey observations are plotted as points with their associated standard error, plotted as vertical lines. This plot was produced using a multi-country set up of the \textit{mcmsupply} functions.}
    \label{fig:fig3_3}
\end{figure}

\subsection{Case 2: Estimating contraceptive method supply shares at the national administration level for a single-country}

\begin{figure}[H]
    \centering
    \includegraphics[width=15cm]{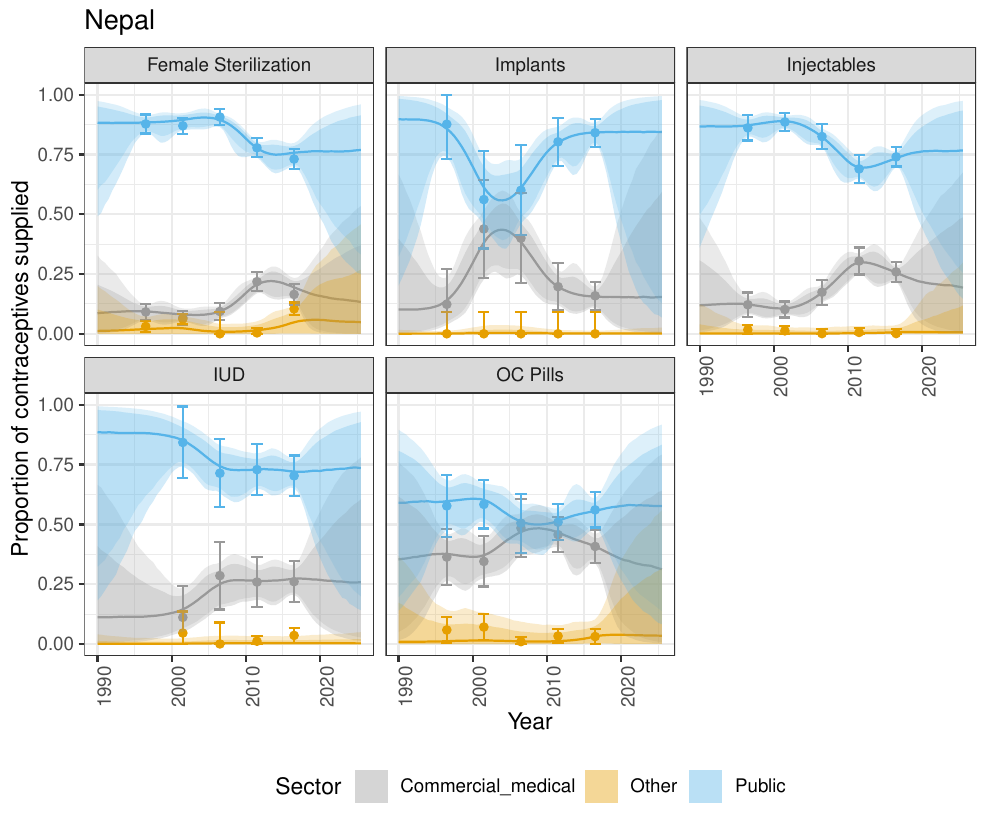}
    \caption{The plotted posterior point estimates for each of the three sectors (public in blue, private commercial medical in grey, and private other in gold) for Nepal at the national administrative level over time with the 80\% and 95\% uncertainty interval denoted as shaded regions. The survey observations are plotted as points with their associated standard error, plotted as vertical lines. This plot was produced using a single-country set up of the \textit{mcmsupply} functions.}
    \label{fig:fig3_4}
\end{figure}

This case considers when the estimates at the national administration level are required for only one country. Rather than running a multi-country model, which takes several hours, a quicker alternative is the single-country approach, which takes only a few minutes. The main difference between the multi-country and single-country model outputs is that the model estimates of the single-country models have slightly larger uncertainty. This is especially evident where data is absent for a particular method. For example in Figure \ref{fig:fig3_3}, the width of the 95\% credible intervals over time for implants estimated by the multi-country national model are smaller than those estimated in the single-country model (Figure \ref{fig:fig3_4}). The arguments \verb|local| and \verb|mycountry| control the single-country estimation models in \textit{mcmsupply}. The user begins as by retrieving the data for Nepal only using the \verb|get_data| function. They set \verb|local=TRUE| and specify which country they are interested in by setting \verb|mycountry=Nepal|.

\begin{verbatim}
cleaned_natdata <- get_data(national=TRUE,  
                            local=TRUE, 
                            mycountry="Nepal")
\end{verbatim}

These arguments are the only discernible differences in the commands for users, the rest of the workflow is as described above in Case 1. A complete workflow for this use case can be found in \verb|vignettes/local_national_mod|. The single-country model produces model estimates that align with those estimated by the multi-country estimation model.

\subsection{Case 3: Estimating contraceptive method supply shares at the subnational administration level for multiple countries simultaneously}

The use case for estimating the contraceptive supply shares via a multi-country model for the subnational administration division is given by the vignette \verb|subnational_multinational _models|. This vignette takes approximately 24 hours to run on a machine with 1.6 GHz Dual-Core Intel Core i5 processor and 8GB of RAM. The dataset contains observations for 225 subnational divisions, across 23 countries. The user begins by calling the multi-country dataset at the subnational administration level via the `national` argument. 

\begin{verbatim}
cleaned_subnatdata <- get_data(national=FALSE)
\end{verbatim}

The remaining workflow is the same as described above in Case 1 and is not shown here. A complete workflow for this use case can be found in \verb|vignettes/subnational_multinational_models|.

\subsection{Case 4: Estimating contraceptive method supply shares at the subnational administration level for a single-country}

This use case is for considering use of the single-country model at the subnational administrative division. The user begins by retrieving the data for Nepal using the \verb|get_data| function by setting the arguments \verb|national=FALSE|, \verb|local=TRUE| and \verb|mycountry="Nepal"|.  

\begin{verbatim}
cleaned_data <- get_data(national=FALSE,  
                         local=TRUE, 
                         mycountry="Nepal")
\end{verbatim}

\begin{figure}[H]
    \centering
    \includegraphics[width=15cm]{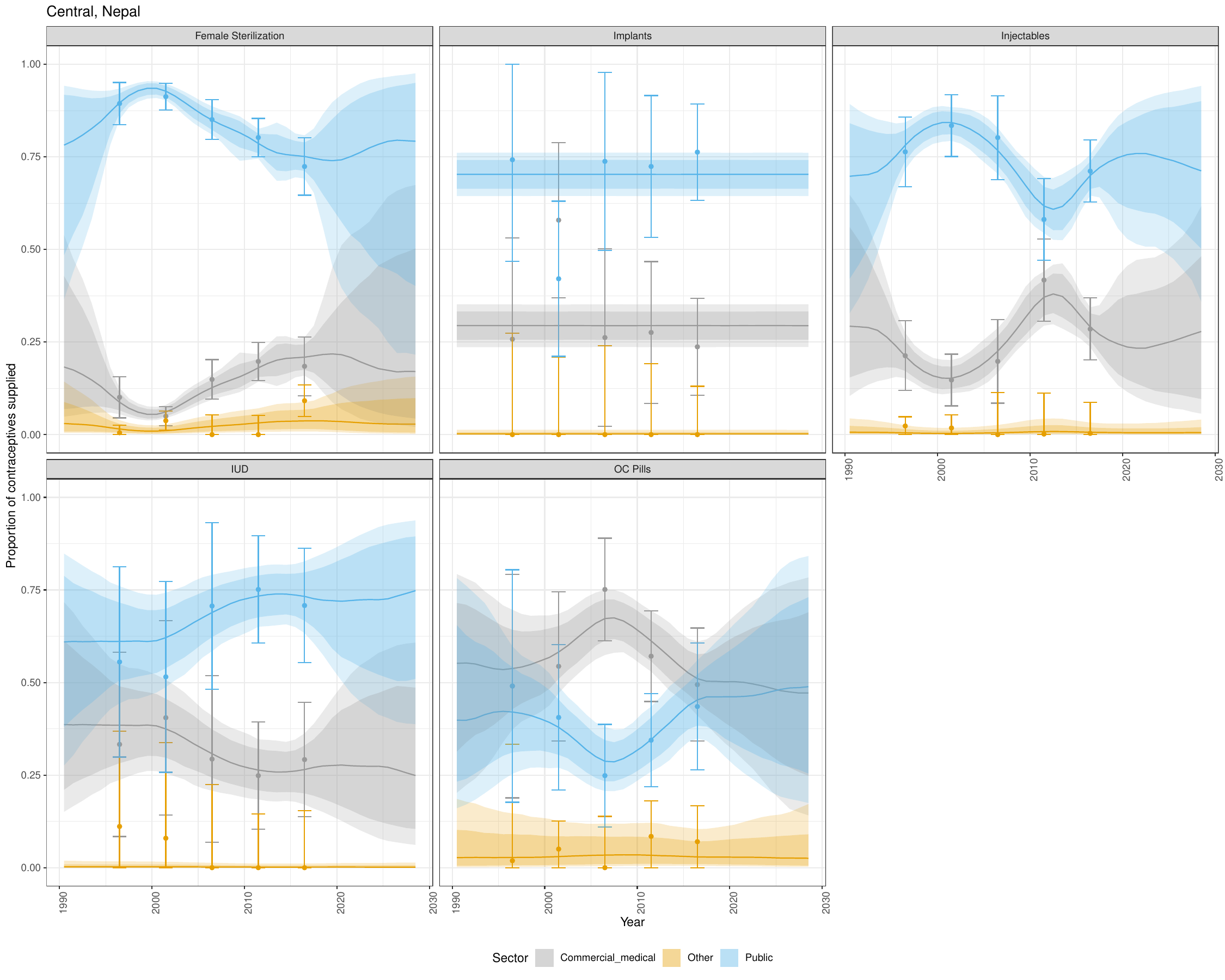}
    \caption{The plotted posterior point estimates for each of the three sectors (public in blue, private commercial medical in grey, and private other in gold) taken from the subnational single-country population for the Central subnational region of Nepal over time with both 80\% and 95\% uncertainty denoted as corresponding shaded regions. The survey observations are plotted as points with their associated standard error, plotted as vertical lines.}
    \label{fig:fig3_5}
\end{figure}

As in the previous use cases, the JAGS model is run and point summaries are calculated via the \verb|run_jags_model| function. The visualisations for the subnational regions of Nepal are returns as a list via the \verb|plot_estimates| function. An example of these visualisations is given in Figure \ref{fig:fig3_5}, where the estimated median method supply shares for Central region of Nepal are plotted with 80\% and 95\% credible intervals over time in each of the methods.

\subsection{Case 5: Estimating contraceptive method supply shares at the national/subnational administration level for a single-country using custom data}

It is possible to include custom datasets when estimating contraceptive method supply shares at either the national or subnational administration level. The set-up for using custom data is very similar to the above processes, with a small difference in the data retrieval step using the \verb|get_data| function. When using a custom dataset, the user defines the location of the `.xlsx` file containing the custom data.

\begin{verbatim}
cleaned_data <- get_data(national=FALSE,  local=TRUE, 
    surveydata_filepath =  "inst/data-raw/my_custom_data_good.xlsx", 
    mycountry="Ethiopia")
\end{verbatim}

The file must be in `.xlsx` format and match the layout of either the \verb|national_FPsource_data| or \verb|subnat_FPsource_data|, depending on your desired administration level. The \verb|get_data| function will carry out a series of internal checks to ensure the custom data matches the stored data layout. If the custom data is not suitable, the \verb|get_data| function will return an error and a message to the user describing the issue with the custom data. Once the custom data checks are complete and passed, the regular workflow for fitting an \textit{mcmsupply} model continues. The JAGS model inputs are retrieved using the \verb|get_modelinputs| and the JAGS model is fit using \verb|run_jags_model|. The summary estimates are plotted using the \verb|plot_estimates| function without any further changes to the workflow. A vignette of how to run the subnational model using a custom dataset can be found \verb|vignettes/subnational_local_customdata_models|.

\section{Discussion} 

In this paper we have introduced the package \textit{mcmsupply}. The primary purpose of this package is to estimate and project modern contraceptive method supply shares from the public, private commercial medical and private other sectors with uncertainty, for a selection of countries participating in the FP2030 initiative. The \textit{mcmsupply} package produces estimates within the period of available Demographic and Health Survey data for a given country, as well as projections beyond the most recent data point. The package uses either stored DHS survey data or custom user-supplied survey data as inputs to the modelling process. The package implements four model types. These models vary by the administration level of their outcome estimates (national or subnational administration level) and dataset type utilised in the estimation (multi-country or single-country contraceptive market supply datasets). The modelling framework uses penalised splines to capture temporal nature of the data. These splines utilise correlations between changes in method supply shares that exist within the data. Using splines informed by cross-method correlations allows us to capture the complex shape of the data without over-fitting. Bayesian hierarchical estimation is another key element of this model estimation process. We take advantage of the geographical nature of the data, such that the expected sector, province (subnational level) or country (national level), method-specific supply shares are informed based on the existing geographical structures in the data. This promotes information sharing across locations, and better inform estimates in areas where smaller amounts of data are present. Case studies illustrate how to use the \textit{mcmsupply} for each of the four potential modelling routes, as well as how to estimate contraceptive method supply shares using custom user-supplied data. \newline

The \textit{mcmsupply} package has many benefits to users. Firstly, it is the first of it's kind to produce annual estimates with uncertainty for monitoring contraceptive method supply shares over time at the national and subnational levels. On average, most countries carry out DHS surveys every 5-6 years, but in some instances the wait time between surveys may be even longer \cite{TheDHSSurveyTypes}. The family planning community use contraceptive method supply shares to evaluate the stability and sustainability of a given country's contraceptive market \cite{Bradley2020}. Contraceptive method supply share estimates are also pivotal in effectively managing contraceptive commodity supply-chains through a 'total market approach' (TMA). A TMA approach to the family planning supply market seeks to engage the public, private commercial, and other sectors in a country to increase family planning users access to vital information, products, and services \cite{Moazzam2015} \cite{SHOPSPlus2016}. Prior to this package, individuals who required contraceptive method supply shares for a given country relied on estimates from the most recent DHS survey in the country, regardless of it's its age. The \textit{mcmsupply} package alleviates this issue and provides the family planning community annual estimates with uncertainty for these contraceptive method supply shares. Secondly, contraceptive method supply shares estimates are not only a stand -alone family planning indicator but are also used in the calculation of an another indicator, estimated modern use (EMU) \cite{Track20EMU}. EMUs aim to measure the proportion of women, aged 15 to 49 years old , who are currently using any modern method of contraception, and are derived from routinely collected family planning service statistics \cite{Track20EMU}. Currently, EMU calculations depend on an adjustment that relies on the most recent DHS survey to provide estimates of the contraceptive supply share market in a given country. Now this adjustment can instead rely on the annual estimates and projections with uncertainty produced by the \textit{mcmsupply} package. This can serve to improve the overall accuracy of EMUs with respect to their ability to accurately measure modern contraceptive use. In addition, given the probabilistic nature of the \textit{mcmsupply} estimates the associated uncertainty can be propagated into the EMU calculations.

The last key benefit of the \textit{mcmsupply} package is the speed at which the user is able to access individual countries supply share estimates. Using the single-country models, at either the national or subnational level, provides users with annual estimates of the method supply shares with uncertainty within minutes. This fast estimation approach is computationally efficient while still producing reliable estimates. Using priors informed by multi-country model subcontinental and country-level median parameter estimates, this modelling approach is robust to spurious parameter estimates even when estimating supply shares for countries with fewer surveys available. This computational efficiency without a loss of model accuracy makes the \textit{mcmsupply} package user-friendly and efficient for regular data analysis. \newline
\newline
The \textit{mcmsupply} is not without its limitations. We remark that the implementation of these models in JAGS have not been optimised. The multi-country models at the national and subnational levels take hours to run and are intensive on computer memory and CPU. Hence, improvements such as matrix operations rather than for-loops would greatly improve the computation efficiency of this package. Another limitation of this package is that in methods without any survey information, the uncertainty intervals tend to be large. This is especially evident in the single-country models where in the absence of data for a given method, the uncertainty of the associated estimates is even larger than that of the corresponding multi-country model estimates. These limitation inspires our future work, where we seek to improve the computational efficiency of these models. We would also like to investigate the potential for incorporating additional covariates into the models, such as average method pricing for each sector, to improve the uncertainty of model estimates and projections where no DHS survey data is available. Lastly, we wish to design an R-shiny app that promotes the use of these model estimates among statistical non-experts within the family planning community.

\section{Conclusions}

\textit{mcmsupply} is an R package that estimates the modern contraceptive method supply shares at the national and subnational administrative divisions over time for countries participating in the Family Planning 2030 initiative. The package provides the user an easy and accessible way to produce annual estimates with uncertainty using Bayesian hierarchical penalised spline models with cross-method correlations at the national and subnational administration levels. These annual estimates with uncertainty may act as a stand-alone family planning indicator of the stability of the modern contraceptive supply market or be used to produce alternative family planning indicators, such as estimated modern use (EMUs) using service statistics \cite{Track20EMU}. To the best of our knowledge, the package is the first of its kind to estimate these supply shares at both administrative levels. Using an R package to disseminate this work aligns with the findability, accessibility, interoperability, and reusability (FAIR) principles of scientific data \cite{Wilkinson2016}. The data used in the package is cited and explained thoroughly, the code is commented and easy to understand should a user wish to tweak or review any functionalities, and finally it is reusable by the very nature of the R package.

\chapter[Model descriptions]{Mathematical description of process to estimate and project contraceptive method supply shares at national and subnational administration levels over time using Bayesian hierarchical penalised splines}
\label{appendix_mcmsupply}

\section{Introduction}
This document is a summary of the extensions made to the model described in Comiskey et al., (2023) \cite{Comiskey2022}.  Described below are the statistical models used to estimate national and subnational contraceptive method supply shares over time using Bayesian hierarchical penalised spline models with multi-country and single-country datasets. These models are utilised in the mcmsupply R package \cite{mcmsupplyR}.

\section{Terminology}

\textit{Contraceptive method supply shares}: The proportion of modern contraceptives supplied by the public, private commercial medical and private other sectors over time. \newline
\newline
\textit{Multi-country national model}: This model estimates the contraceptive method supply shares at the national administration level over time for many countries simultaneously.\newline
\newline
\textit{Single-country national model}: This model estimates the contraceptive method supply shares at the national administration level over time for a single country.\newline
\newline
\textit{Multi-country subnational model}: This model estimates the contraceptive method supply shares at the subnational administration level over time for many countries simultaneously.\newline
\newline
\textit{Single-country subnational model}: This model estimates the contraceptive method supply shares at the subnational administration level over time for a single country.\newline

\section{Overall model set-up}
The outcome of interest is the components of a compositional vector 
$$\boldsymbol{{\phi_{q,t,m}}} =(\phi_{q,t,m,s=1},\phi_{q,t,m,s=2},\phi_{q,t,m,s=3})$$ 
where, $\phi_{q,t,m,s}$ is the proportion supplied by the public sector (s =1), the private commercial medical sector (s=2) and the other private sector (s=3) of modern contraceptive method m, at time t, in population q (national or subnational). \newline 
\newline
We begin by defining a regression model for $\phi_{q,t,m,1}$.  The logit-transformed proportion, logit($\phi_{q,t,m,1}$), is modelled through a latent variable $\psi_{q,t,m,1}$, with a penalized basis-spline (P-spline) regression model:

\begin{equation}
    \operatorname{logit}\left(\phi_{q,t,m,1}\right)=\psi_{q,t,m,1}=\sum_{k=1}^K \beta_{q,m,1,k} B_{q,k}(t),
\end{equation}
 
where, \newline
$\psi_{q,t,m, 1}$ is the latent variable capturing the logit proportions of the public sector (s=1) supply share of method m, at time t, in population q. $B_{q,k}(t)$ refers to the $k^{th}$ basis function evaluated in population q, at time t. $\beta_{q,m,1,k}$ is the $k^{th}$ spline coefficient for the public sector supply (s=1) of method m in population q. \newline

Similarly, we model the latent variable, $\psi_{q,t,m,2}$, to capture the logit-transformed ratio of the private commercial medical supply share to the total private sector share. The model is specified as follows:

\begin{equation}
\begin{aligned}
    \operatorname{logit}\left(\frac{\phi_{q,t, m, 2}}{1 - \phi_{q,t, m, 1}}\right)=\psi_{q, t, m, 2}=\sum_{k=1}^K \beta_{q, m, 2, k} B_{q,k}(t),
\end{aligned}
\end{equation}

where, $\beta_{q,m,2,k}$ is the $k^{th}$ spline coefficient for the ratio of private commercial medical sector (s=2) to total private sector for method m in population q. \newline

The basis functions $B_{k}(t)$ are constructed using cubic splines. The basis are fitted over the years 1990 to 2025. We align the knot placement of the basis splines with the most recent survey year in each country. As the most recent survey year varies by country (in the case of national-level data) or province (in the case of subnational data), the basis splines $B_{q,k}(t)$ also vary by location.\newline

To estimate the spline coefficients, $\beta_{q,m,s,k}$  we model them indirectly by estimating the penalised first order differences of the spline coefficients, $\boldsymbol{\delta_{q,m,s}}$. This reparametrization of the spline coefficients ensures that the model estimates are smooth when projecting into the future. The  $\boldsymbol{\delta_{q,m,s}}$  vector is of length h where h=K-1, and K is the total number of knots used in the set of basis functions. It is defined as,
\begin{equation}
\begin{aligned}
    \boldsymbol{\delta_{q,m,s}} = (\beta_{q,m, s, 2} - \beta_{q,m, s, 1}, \beta_{q,m, s, 3} - \beta_{ m, s, 2}, ...., \beta_{ m, s, K} - \beta_{q,m, s, K-1}).
\end{aligned}
\end{equation}

We assume that in population q, for method m and sector s, the value of spline coefficient at knot index k*, aligning with the year t*, the most recent survey available, is $\alpha_{q, m, s}$. By doing this, we are assuming that the $\alpha_{q,m,s}$ parameter will act as the spline coefficient for the reference spline at k*. We are then able to calculate the remaining spline coefficients from the reference index (k*) using the estimated $\boldsymbol{\delta_{q,m,s}}$.

\begin{equation}
    \beta_{q,m, s, k}=\left\{\begin{array}{c}
    \alpha_{q,m,s} \quad k=k^*, \\
    \beta_{q,m,s, k+1}-\delta_{q, m, s, k} \quad k<k^*, \\
    \beta_{q,m,s, k-1}+\delta_{q, m, s, k-1} \quad k>k^*
\end{array}\right.
\end{equation} 
Where, \newline
$\alpha_{q,m,s}$ is the most recently observed supply share on the logit scale for sector s , method m, in population q. This proxies as an intercept in the model. \newline
k is the knot index along the set of basis splines $B_{q,k}(t)$ \newline
$k^*$ is the index of the knot that corresponds with $t^*$, the year index where the most recent survey occurred in population q. \newline
$\delta_{q,m,s,k-1}$ is the first order difference between spline coefficients $\beta_{q,m, s, K}$ and $\beta_{q,m, s,K-1}$ \newline

We assume a smooth transition between spline coefficients. Thus, we centre our rates of change, $\delta_{q,1:M,s,k}$ , on 0, with a variance-covariance matrix, $\Sigma_{\delta_{s}}$, that captures the correlations that exist between the rates of change in supply shares for each pair of methods.

\begin{equation}
    \delta_{q,1:M,s,h} \mid \Sigma_{\delta_{s}}  \sim MVN(\boldsymbol{0}, \Sigma_{\delta_{s}}),
\end{equation}

From the latent variable vector,  $\boldsymbol{\psi_{q,t,m}}$, it is possible to infer the compositional vector $\boldsymbol{{\phi_{q,t,m}}}$,

\begin{equation}
\begin{aligned}
&\phi_{q,t, m, 1}=\operatorname{logit}^{-1}\left(\psi_{q,t, m, 1}\right),\\
&\phi_{q,t, m, 2}=\left(1-\phi_{q,t, m, 1}\right) \operatorname{logit}^{-1}\left(\psi_{q,t, m, 2}\right),\\
&\phi_{q,t, m, 3}=1-\left(\phi_{q,t, m, 1}+\phi_{q,t, m, 2}\right) .
\end{aligned}
\end{equation}

The likelihood of the logit-transformed observed data, $\operatorname{logit}(y_{i})$, the observed logit-transformed proportion of modern contraceptive method supplied by the public and commercial medical sectors (s=1 or s=2) for method m, at time t are modelled using Normal distributions such that

\begin{equation}
    \operatorname{logit}(y_{i}) \mid \psi_{q[i],t[i], m[i], s[i]}  \sim N(\psi_{q[i],t[i], m[i], s[i]}, SE_{i}^2).
\end{equation}
Where,\newline
$\psi_{q[i], t[i], m[i], s[i]}$ is the logit-transformed supply proportion for the population q, time t, method m, and sector s associated with observation i. \newline
The logit-transformed variance, $SE_{i}^2$, utilizes the standard error (SE) calculated using the DHS survey microdata associated with observation $y_{i}$. The variance is transformed onto the logit scale using the delta-method \cite{Casella2002}.

\section{Estimation of model parameters}

\begin{landscape}
\begin{figure}
    \centering
    \includegraphics[width=20cm]{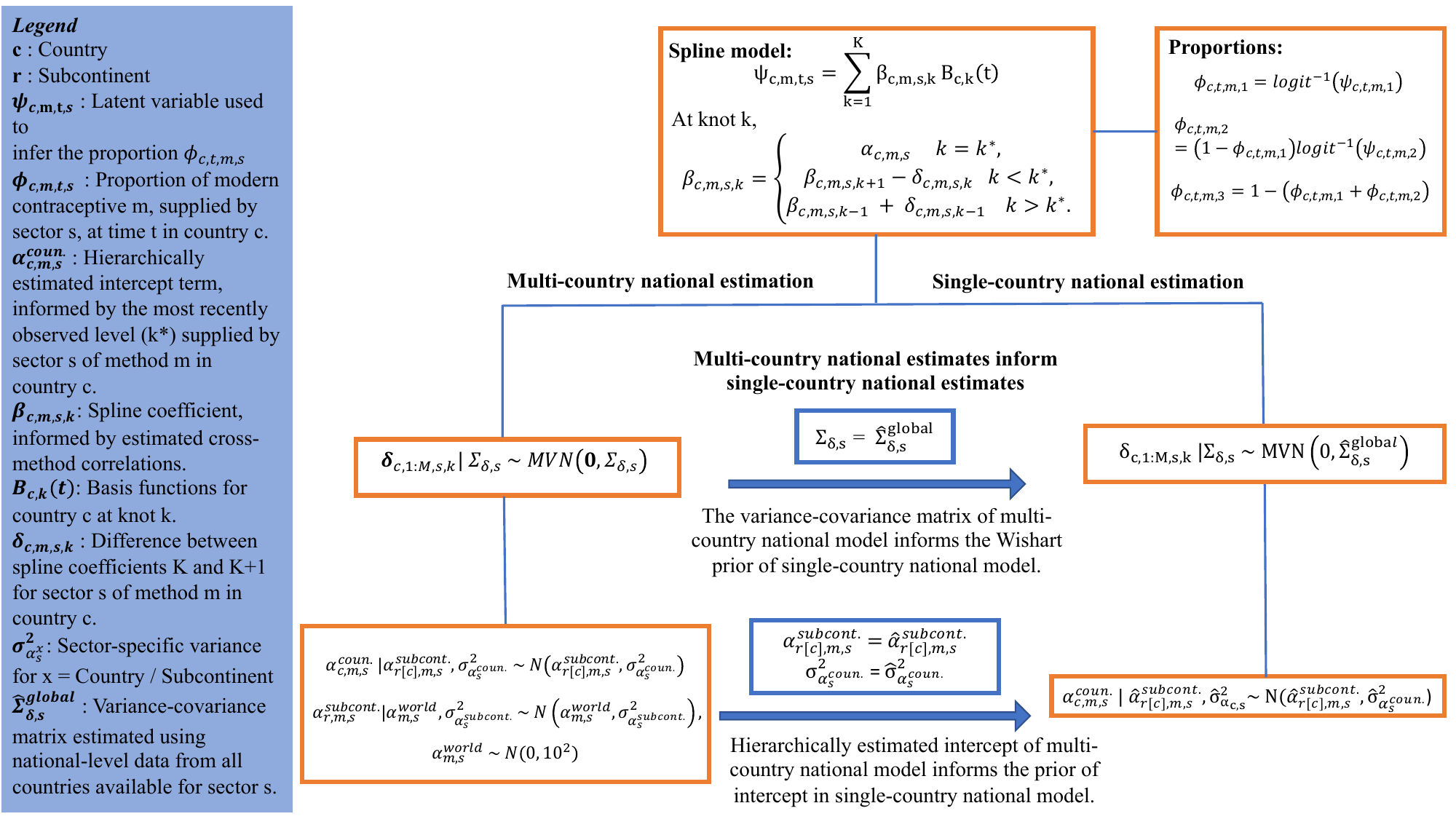}
    \caption{Schematic linking the multi-country and single-country national-level modelling approaches.}
    \label{fig:nat_modelling_scheme}
\end{figure}
\end{landscape}

\begin{landscape}
\begin{figure}
    \centering
    \includegraphics[width=20cm]{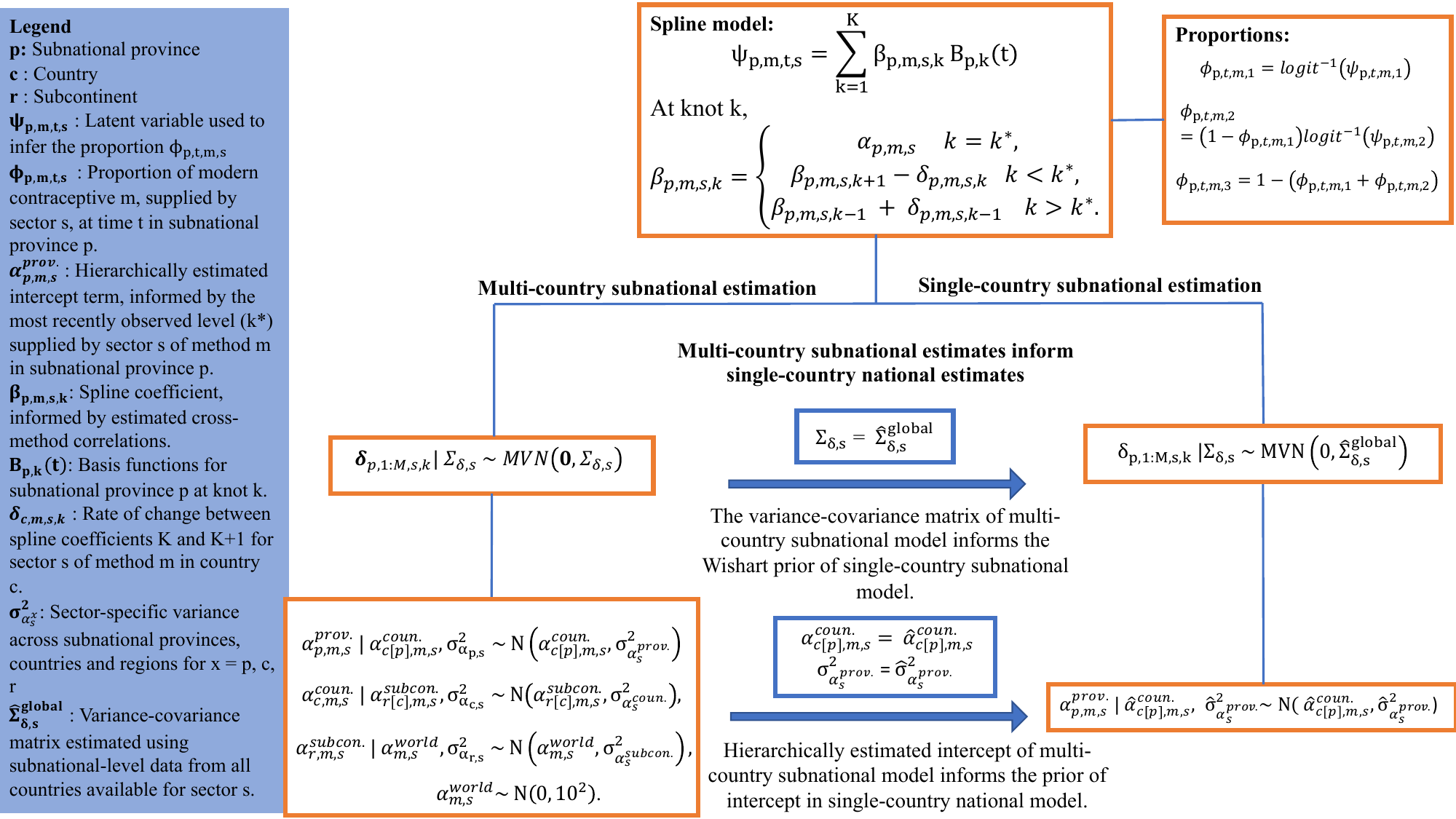}
    \caption{Schematic linking the multi-country and single-country subnational-level modelling approaches.}
    \label{fig:subnat_modelling_scheme}  
\end{figure}
\end{landscape}

Summaries of the national and subnational level models can be found in Figure \ref{fig:nat_modelling_scheme} and Figure \ref{fig:subnat_modelling_scheme}. A table of parameters and their interpretations can be found in Table \ref{tab:param_table}.

\subsection{Modelling \texorpdfstring{$\alpha_{q,m,s}$} \textbf{hierarchically with a multi-country dataset}}
In this approach, we take advantage of the geographic nature of the dataset. We pool data to estimate precise intercepts at higher geographic levels that then go on to inform more granular level intercepts, until we reach our geographic level of interest (national or subnational), where less data is present. 

\subsubsection{National level model}
At the national level, the hierarchical distributions to capture the most recently observed DHS level in country c, for method m, supplied by sector s, are given by:
\begin{equation}
\begin{aligned}
\alpha_{c, m, s}^{country} \mid \theta_{r[c], m, s}^{subcon.}, \sigma_{\alpha, s}^2 & \sim N\left(\theta_{r[c], m, s}^{subcon.}, \sigma_{\alpha, s}^2\right), \\
\theta_{r, m, s}^{subcon.} \mid \theta_{w, m, s}^{world}, \sigma_{\theta, s}^2 & \sim N\left(\theta_{w, m, s}^{world}, \sigma_{\theta, s}^2\right), \\
\theta_{w, m, s}^{world} \sim N\left(0, 10^2\right), \\
\sigma_{\alpha, s} \sim Cauchy\left(0, 1\right)_{+}, \\
\sigma_{\theta, s} \sim Cauchy\left(0, 1\right)_{+}.
\end{aligned}
\end{equation}
Where, the geographic hierarchy begins at the world level $\theta_{w, m, s}^{world}$, which informs the subcontinental intercepts, $\theta_{r, m, s}^{subcon.}$, which in turn inform individual country intercepts, $\alpha_{c, m, s}^{country}$. Vaguely informative Cauchy priors are given to the standard deviation terms of the country- and subcontinental- terms \cite{Gelman2006PriorDraper}. The standard deviation terms capture the cross-country ($\sigma_{\alpha, s}$) and cross-subcontinent ($\sigma_{\theta, s}$) variation within the data.

\subsubsection{Subnational level model}
At the subnational level, we include an additional layer of geographic intercepts to capture the most recently observed DHS level in subnational province p, for method m,supplied by sector s. While, we use the same notation to explain the hierarchical set up of this approach, the estimates of the country-level and above parameters will be different from the national-level model to the subnational-level model. In the subnational instance, the hierarchical distributions are given by:

\begin{equation}
\begin{aligned}
\alpha_{p, m, s}^{prov.} \mid \alpha_{c[p], m, s}^{country}, \sigma_{\alpha_{p}, s}^2 & \sim N\left(\alpha_{c[p], m, s}^{country}, \sigma_{\alpha_{p},s}^2\right), \\
\alpha_{c, m, s}^{country} \mid \theta_{r[c], m, s}^{subcon.}, \sigma_{\alpha_{c}, s}^2 & \sim N\left(\theta_{r[c], m, s}^{subcon.}, \sigma_{\alpha_{c},s}^2\right), \\
\theta_{r, m, s}^{subcon.} \mid \theta_{w, m, s}^{world}, \sigma_{\theta, s}^2 & \sim N\left(\theta_{w, m, s}^{world}, \sigma_{\theta, s}^2\right), \\
\theta_{w, m, s}^{world} \sim N\left(0, 10^2\right), \\
\sigma_{\alpha_{p},s} \sim Cauchy\left(0, 1\right)_{+}, \\
\sigma_{\alpha_{c},s} \sim Cauchy\left(0, 1\right)_{+}, \\
\sigma_{\theta_{s}} \sim Cauchy\left(0, 1\right)_{+}.  
\end{aligned}
\end{equation}

In this instance, we mirror the geographic hierarchy of the national model, and add an additional layer to reflect the province-level intercepts, $\alpha_{p, m, s}^{prov.}$, of the subnational level model, and cross-provincial variation ($\sigma_{\alpha_{p}, s}$).

\subsection{Modelling \texorpdfstring{$\alpha_{q,m,s}$} \textbf{ using informative priors with a single-country dataset}}
In this approach, priors for higher-population level intercept parameters are informed from the multi-country national or subnational level models (i.e., the models that used multi-country datasets). 

\subsubsection{National-level model}
$\alpha_{c,m,s}^{country}$  is the national-level intercept for country c, method m and sector s, informed by the posterior median estimates of the subcontinental level model interceptand the associated variance parameter estimated from the multi-country national model, such that 

\begin{equation}
\alpha_{c, m, s}^{country} \mid \hat{\theta}_{r[c], m, s}^{subcon.}, \hat{\sigma}_{\alpha_{c},s}^2 \sim N\left(\hat{\theta}_{r[c], m, s}^{subcon.}, \hat{\sigma}_{\alpha_{c},s}\right),
\end{equation} 
where, $\hat{\theta}_{r[c], m, s}^{subcon.}$ is the posterior median UNSD subcontinental population intercept for region r, method m, sector s, associated with country c estimated from the national-level multi-country model and 
$\hat{\sigma}_{\alpha_{c},s}^2$ is the posterior median of the sector specific cross-country variation associated with the $\alpha_{c,m,s}$ intercept estimated from the national-level multi-country model.\newline

\subsubsection{Subnational-level model}

$\alpha_{p,m,s}^{prov.}$  is the subnational-level intercept for subnational province p, method m, sector s,  informed by the posterior median estimates of the country-level model interceptand the associated  variance parameter estimated from the multi-country subnational model, such that 

\begin{equation}
\alpha_{p, m, s}^{prov.} \mid \hat{\alpha}_{c, m, s}^{country}, \hat{\sigma}_{\alpha_{p},s}^2 \sim N\left(\hat{\alpha}_{c, m, s}^{country}, \hat{\sigma}_{\alpha_{p},s}^2\right),
\end{equation}

where, $\hat{\alpha}_{c[p], m, s}^{country}$ is the posterior median  national-level population intercept for country c, method m, sector s, associated with the subnational province p, estimated from the subnational-level multi-country model and 
$\hat{\sigma}_{\alpha_{p},s}^2$ is the posterior median of the cross-province variation associated with the $\alpha_{p,m,s}$ intercept estimated from the subnational-level multi-country model.

\subsection{Modelling \texorpdfstring{$\Sigma_{\delta_{s}}$} \textbf{using cross-method correlations with a multi-country dataset}}

In this approach, we decompose the $\Sigma_{\delta_{s}}$ into its variance and correlation matrices and estimate the components separately. This is a two-model run approach which involves estimating the correlations using a model run with correlations set to 0. 

For both the national and subnational models, a multivariate normal prior centred on 0 was assigned to the vector of length M of first-order differences of the spline coefficients, $\delta_{q,1:M,s,h}$, for population q (national or subnational), using all methods supplied by sector s at first-order difference h;

\begin{equation}
    \delta_{q,1:M,s,h} \mid \Sigma_{\delta_{s}}  \sim MVN(\boldsymbol{0}, \Sigma_{\delta_{s}}),
\end{equation} 
where,
\begin{equation}
    \Sigma_{\delta_s}=\left[\begin{array}{ccccc}
    \sigma_{\delta_{1, s}}^2 & \hat{\rho}_{1,2, S} \sigma_{\delta_{1, S}} \sigma_{\delta_{2, s}} & \ldots & \ldots & \hat{\rho}_{1, M, S} \sigma_{\delta_{1, s}} \sigma_{\delta_{M, S}} \\
    \hat{\rho}_{2,1, s} \sigma_{\delta_{2, s}} \sigma_{\delta_{1, s}} & \sigma_{\delta_{2, s}}^2 & \ldots & \ldots & \hat{\rho}_{2, M, S} \sigma_{\delta_{2, s}} \sigma_{\delta_{M, s}} \\
    \ldots & \ldots & \ldots & \ldots & \ldots \\
    \ldots & \ldots & \ldots & \ldots & \ldots \\
    \hat{\rho}_{M,1, s} \sigma_{\delta_{M, s}} \sigma_{\delta_{M, s}} & \ldots & \ldots & \ldots & \sigma_{\delta_{M, s}}^2
\end{array}\right].
\end{equation}
The correlation terms of the covariance matrix, $\rho_{i,j,s}$, were estimated using a maximum a posteriori estimator for the correlation matrix as described in Azose and Raftery, 2018 \cite{Azose2018}. This approach involves fitting a model where the covariance terms in $\sigma_{\delta_{j, s}}$ are set equal to zero. The national and subnational models deviate in terms of the geographic level that they estimate these correlations at. For the national model, we estimate the correlations based on country-level rates of change, $\delta_{c,1:M,s,h}$, terms. This captures the correlations across all countries. While in the subnational model, we estimate the correlations using provincial-level rates of change, $\delta_{p,1:M,s,h}$. In this instance, the correlations captured are across the subnational provinces of all countries.

\subsubsection{National model}

In the national model, the deviation terms of the $\Sigma_{\delta_s}$ matrix are given vague uniform priors,

\begin{equation}
\begin{aligned}
    \sigma_{\delta_{m,s}} \sim Uniform\left(0, 10\right).
\end{aligned}
\end{equation}
The 0-covariance model estimates are used to estimate the correlation between methods across time and all countries at the national-level. Specifically, for sector s, the correlation between method i and method j is calculated as follows,

\begin{equation}
\hat{\rho}_{i,j,s} = \frac{\sum_{c=1}^C \sum_{h=1}^{K-1}\tilde{\delta}_{c,m[i],s,h}\tilde{\delta}_{c,m[j],s,h}}{\sqrt{\sum_{c=1}^C \sum_{h=1}^{K-1}\tilde{\delta}_{c,m[i],s,h}^2} \sqrt{\sum_{c=1}^C \sum_{h=1}^{K-1}\tilde{\delta}_{c,m[j],s,h}^2}}.
\end{equation} 
Where, $\tilde{\delta}_{c,m[j],s,h}$  are the estimated first order differences of the spline coefficients for country c, method m, sector s, at the h-th difference between spline coefficients. They are given by the posterior medians of $\delta_{c,m,s,h}$ from the zero-covariance run, after subsetting the period considered to periods with data within a country. 
C represents the total number of countries involved in the study
K is the number of knots in the basis functions 
h represents the number of differences (h=K-1) between the spline coefficients. \newline

\subsubsection{Subnational model}

In the subnational model, the deviation terms of the $\Sigma_{\delta_{s}}$ matrix are given vague Cauchy priors. This prior is suggested as a weakly informative prior in the paper titled Prior distributions for variance parameters in hierarchical models by Gelman, Bayesian Analysis (2006) \cite{Gelman2006PriorDraper}. 

\begin{equation}
\begin{aligned}
    \sigma_{\delta_{m,s}} \sim Cauchy\left(0, 1\right)_{+}.
\end{aligned}
\end{equation}
The 0-covariance model estimates are used to estimate the strength of the correlations between methods across time and all provinces in all countries at the subnational-level. Specifically, for sector s, the correlation between method i and method j is calculated as follows,

\begin{equation}
\hat{\rho}_{i,j,s} = \frac{\sum_{p=1}^P \sum_{h=1}^{K-1}\tilde{\delta}_{p,m[i],s,h}\tilde{\delta}_{p,m[j],s,h}}{\sqrt{\sum_{p=1}^P \sum_{h=1}^{K-1}\tilde{\delta}_{p,m[i],s,h}^2} \sqrt{\sum_{p=1}^P \sum_{h=1}^{K-1}\tilde{\delta}_{p,m[j],s,h}^2}}
\end{equation} \newline
Where, $\tilde{\delta}_{p,m[j],s,h}$  are the estimated first order differences of the spline coefficients for province p, method m, sector s, at the h-th difference between spline coefficients. They are given by the posterior medians of $\delta_{p,m,s,h}$ from the zero-covariance run, after subsetting the period considered to periods with data within each province. \newline
P represents the total number of subnational provinces across all countries involved in the study. K is the number of knots in the basis functions. h represents the number of differences (h=K-1) between the spline coefficients.

\subsection{Modelling \texorpdfstring{$\Sigma_{\delta_{s}}$} \textbf{using informative priors with a single-country dataset}}
For estimation of the method supply shares using a single-country national or subnational dataset, we set $\Sigma_{\delta_{s}}$ as the median estimate of the MxM variance-covariance matrix from the corresponding (national or subnational) multi-country model, $\Sigma_{\delta_s}^{\textbf{global}}$ and we estimate the first-order difference spline coefficients using a Multivariate Normal prior centred on zero such that
\begin{equation}
    \delta_{1:M,s,h} \mid \hat{\Sigma}_{\delta_s}^{\textbf{global}}  \sim MVN(\boldsymbol{0}, \hat{\Sigma}_{\delta_s}^{\textbf{global}}).
\end{equation}

\begin{table}[ht]
\resizebox{\textwidth}{!}{%
\begin{tabular}{|c|c|}
\hline
\textbf{Model Parameter} &
  \textbf{Interpretation} \\ \hline
$\phi_{q,t,m,s}$ &
  \begin{tabular}[c]{@{}c@{}}The proportion supplied by sector s, of modern contraceptive method m, \\ at time t,  in population q \\ (national or subnational).\end{tabular} \\ \hline
$\psi_{q,t,m,s}$ &
  The latent variable used to model $\phi_{q,t,m,s}$ on the logit scale \\ \hline
$\beta_{q,m,s,k}$ &
  The $k^{th}$ spline coefficient for sector s, method m in population q \\ \hline
$B_{q,k}(t)$ &
  The set of basis functions for population q, evaluated at knot k for time t. \\ \hline
$\alpha_{q,m,s}^{pop.}$ &
  \begin{tabular}[c]{@{}c@{}}The most recently observed supply share on the logit scale \\ for sector s , method m, in population q. \\ This proxies as an intercept in the model.\end{tabular} \\ \hline
$\delta_{q,m,s,k-1}$ &
  \begin{tabular}[c]{@{}c@{}}The first order difference between spline coefficients \\ $\beta_{q,m, s, K}$ and $\beta_{q,m, s,K-1}$\end{tabular} \\ \hline
$\Sigma_{\delta_{s}}$ &
  Variance-covariance matrix used in the MVN prior of $\delta_{q,1:M,s,h}$ \\ \hline
$\sigma_{X,s}$ &
  \begin{tabular}[c]{@{}c@{}}Standard deviation terms relating to the intercept \\ parameter X for sector s.\\ X may be at the provincial-, country-, or subcontinental-level.\end{tabular} \\ \hline
$\rho_{i,j,s}$ &
  \begin{tabular}[c]{@{}c@{}}Correlation between the rates of change in supply shares \\ for method{[}i{]} and method{[}j{]} in sector s.\end{tabular} \\ \hline
\end{tabular}%
}
\caption{A table of parameters names and their interpretations across the national and subnational models. The indexing refers to sector s, of modern contraceptive method m, at time t,  in population q (national or subnational).}
\label{tab:param_table}
\end{table}

\chapter[Model validations]{Validation of the multi-country and single-country subnational models and the single-country national model} 

\section{Data}
\subsection{National data source}
In this study we consider countries involved in the FP2030 initiative. A database of the public and private sector breakdown of modern contraceptive supply with their associated standard errors at the national administration level was created using data from the DHS \cite{DHS_survey_methods}.  Table \ref{tab:table_A1} lists the thirty countries used in the national level method supply share database. The total number of surveys carried out and the year of the most recent survey is listed for each country. A full description of the national level data can be found in Comiskey et al., 2023 \cite{Comiskey2022}.

\begin{table}[H]
\resizebox{\textwidth}{!}{%
\begin{tabular}{|c|c|c|c|}
\hline
\textbf{\begin{tabular}[c]{@{}c@{}}UNSD intermediate \\ world regions\end{tabular}} &
  \textbf{Country} &
  \textbf{\begin{tabular}[c]{@{}c@{}}Total Number \\ of Surveys\end{tabular}} &
  \textbf{Recent Survey Year} \\ \hline
Southern Asia      & Afghanistan                  & 1  & 2015 \\ \hline
Western Africa     & Benin                        & 5  & 2017 \\ \hline
Western Africa     & Burkina Faso                 & 4  & 2010 \\ \hline
Middle Africa      & Cameroon                     & 5  & 2018 \\ \hline
Middle Africa      & Congo                        & 1  & 2005 \\ \hline
Middle Africa      & Democratic Republic of Congo & 2  & 2013 \\ \hline
Western Africa     & Cote d'Ivoire                & 3  & 2011 \\ \hline
Eastern Africa     & Ethiopia                     & 5  & 2019 \\ \hline
Western Africa     & Ghana                        & 5  & 2014 \\ \hline
Western Africa     & Guinea                       & 4  & 2018 \\ \hline
Southern Asia      & India                        & 4  & 2005 \\ \hline
Eastern Africa     & Kenya                        & 5  & 2014 \\ \hline
Western Africa     & Liberia                      & 4  & 2019 \\ \hline
Eastern Africa     & Madagascar                   & 4  & 2008 \\ \hline
Eastern Africa     & Malawi                       & 5  & 2015 \\ \hline
Western Africa     & Mali                         & 5  & 2018 \\ \hline
Eastern Africa     & Mozambique                   & 3  & 2011 \\ \hline
South-Eastern Asia & Myanmar                      & 1  & 2015 \\ \hline
Southern Asia      & Nepal                        & 5  & 2016 \\ \hline
Western Africa     & Niger                        & 4  & 2012 \\ \hline
Western Africa     & Nigeria                      & 5  & 2018 \\ \hline
Southern Asia      & Pakistan                     & 4  & 2017 \\ \hline
South-Eastern Asia & Philippines                  & 6  & 2017 \\ \hline
Eastern Africa     & Rwanda                       & 6  & 2019 \\ \hline
Western Africa     & Senegal                      & 10 & 2019 \\ \hline
Western Africa     & Sierra Leone                 & 3  & 2019 \\ \hline
Eastern Africa     & Tanzania                     & 6  & 2015 \\ \hline
Western Africa     & Togo                         & 2  & 2013 \\ \hline
Eastern Africa     & Uganda                       & 5  & 2016 \\ \hline
Eastern Africa     & Zimbabwe                     & 5  & 2015 \\ \hline
\end{tabular}%
}
\caption{Summary of DHS microdata used for national level estimation. This table provides the United Nation Statistics Division (UNSD) intermediate world region names, country names, the number of DHS surveys per country available and the year of the most recent DHS survey available. Just over 46\% of countries have data available after 2015.}
\label{tab:table_A1}
\end{table}

\subsection{Subnational data source}
In this study we consider countries involved in the FP2030 initiative. A database of administration-1 level Demographic and Health Survey (DHS) data observations for the supply of modern contraceptive methods by the public and private sectors and their associated standard errors was created using the Integrated Public Use Microdata Series (IPUMS) project, IPUMS-DHS \cite{IPUMSdata}.  Like the national-level study \cite{Comiskey2022}, the modern methods of contraception considered in this study are female sterilisation, oral contraceptive pills (OC pills), implants (including Implanon, Jadelle and Sino-implant), intra-uterine devices (IUD, including Copper- T 380-A IUD and LNG-IUS), and injectables (including Depo Provera (DMPA), Noristerat (NET-En), Lunelle, Sayana Press and other injectables).  The variables contained within the IPUMS-DHS database are consistent over time and space.  IPUMS-DHS uses integrated geography variables for a country across sample years to address issues with subnational boundaries changing over time and enable comparisons over time. Table \ref{tab:tab_6_1} lists the 23 countries captured in this subnational database. 
The total number of administration level 1 (admin-1) subnational regions, the number of DHS surveys each country has in the database, and the year of the most recent survey in the database is listed for each country. Just under half of the countries included have survey data available after 2015, highlighting the need for annual up-to-date estimates of the contraceptive supply shares.

\begin{table}[ht]
\resizebox{\textwidth}{!}{%
\begin{tabular}{|c|c|c|c|}
\hline
\textbf{Country} &
  \textbf{\begin{tabular}[c]{@{}c@{}}Number of admin-1 level \\ subnational provinces\end{tabular}} &
  \textbf{\begin{tabular}[c]{@{}c@{}}Number of IPUMS-DHS \\ surveys\end{tabular}} &
  \textbf{Recent survey year} \\ \hline
Benin                     & 6  & 4 & 2017 \\ \hline
Burkina Faso              & 13 & 4 & 2010 \\ \hline
Cameroon                  & 3  & 3 & 2004 \\ \hline
Congo Democratic Republic & 5  & 2 & 2013 \\ \hline
Cote d'Ivoire             & 15 & 3 & 2011 \\ \hline
Ethiopia                  & 10 & 4 & 2016 \\ \hline
Ghana                     & 8  & 5 & 2014 \\ \hline
Guinea                    & 3  & 4 & 2018 \\ \hline
India                     & 27 & 4 & 2015 \\ \hline
Kenya                     & 8  & 5 & 2014 \\ \hline
Liberia                   & 5  & 2 & 2013 \\ \hline
Madagascar                & 6  & 4 & 2008 \\ \hline
Malawi                    & 3  & 5 & 2016 \\ \hline
Mali                      & 4  & 5 & 2018 \\ \hline
Mozambique                & 11 & 3 & 2011 \\ \hline
Nepal                     & 5  & 5 & 2016 \\ \hline
Niger                     & 6  & 4 & 2012 \\ \hline
Pakistan                  & 6  & 4 & 2017 \\ \hline
Rwanda                    & 7  & 6 & 2014 \\ \hline
Senegal                   & 4  & 9 & 2017 \\ \hline
Tanzania                  & 6  & 6 & 2015 \\ \hline
Uganda                    & 4  & 4 & 2016 \\ \hline
Zimbabwe                  & 10 & 5 & 2015 \\ \hline
\end{tabular}%
}
\caption{Summary information regarding the countries considered for subnational modelling. The name, number of subnational administration level 1 (admin-1) regions, the total number of DHS surveys present in the data, and the year of the most recent DHS survey in the data for each country are listed.}
\label{tab:tab_6_1}
\end{table}

\subsection{Standard error calculation}

Both the national-level and subnational-level method supply share databases use the ‘svyciprop’ function from the ‘survey’ package in R to calculate the standard errors associated with each observation \cite{Lumley2021}. Using DHS design factors, we impute calculated standard errors when the calculated standard error is 0. A full description of how the standard errors are calculated and the imputation technique used to estimate standard errors can be found in the supplementary material of Comiskey et al., (2023) sections 3.1 and 3.2. 

\begin{table}[ht]
\resizebox{\textwidth}{!}{%
\begin{tabular}{|c|c|c|cc|cc|}
\hline
\textbf{Measure} &
  \textbf{\begin{tabular}[c]{@{}c@{}}Range \\ (\% over all methods)\end{tabular}} &
  \textbf{\begin{tabular}[c]{@{}c@{}}Median SE size \\ (\% over all methods)\end{tabular}} &
  \multicolumn{2}{c|}{\textbf{\begin{tabular}[c]{@{}c@{}}Largest mean SE\\ (method, \%)\end{tabular}}} &
  \multicolumn{2}{c|}{\textbf{\begin{tabular}[c]{@{}c@{}}Smallest mean SE\\ (method, \%)\end{tabular}}} \\ \hline
Result &
  0.015 , 18.19 &
  2.23 &
  \multicolumn{1}{c|}{IUD} &
  4.10 &
  \multicolumn{1}{c|}{Injectables} &
  2.03 \\ \hline
\end{tabular}%
}
\caption{Summary table for the calculated standard errors of the national DHS data observations.}
\label{tab:my-natSEtable}
\end{table}

From Table \ref{tab:my-natSEtable}, the calculated national-level standard errors range from 0.015 to 18.19 percentage points. The median standard error size across all method is 2.23 percentage points. On average, they tend to be largest for IUDs where the mean standard error size is approximately 4 percentage points and smallest in injectables where the mean standard error size is approximately 2 percentage points. \newline

\begin{table}[H]
\resizebox{\textwidth}{!}{%
\begin{tabular}{|c|c|c|cc|cc|}
\hline
\textbf{Measure} &
  \textbf{\begin{tabular}[c]{@{}c@{}}Range \\ (\% over all methods)\end{tabular}} &
  \textbf{\begin{tabular}[c]{@{}c@{}}Median SE size \\ (\% over all methods)\end{tabular}} &
  \multicolumn{2}{c|}{\textbf{\begin{tabular}[c]{@{}c@{}}Largest mean SE\\ (method, \%)\end{tabular}}} &
  \multicolumn{2}{c|}{\textbf{\begin{tabular}[c]{@{}c@{}}Smallest mean SE\\ (method, \%)\end{tabular}}} \\ \hline
Result &
  0.0 , 22.0 &
  3.84 &
  \multicolumn{1}{c|}{OC pills} &
  5.3 &
  \multicolumn{1}{c|}{Implants} &
  3.6 \\ \hline
\end{tabular}%
}
\caption{Summary table for the calculated standard errors of the subnational IPUMS-DHS data observations. }
\label{tab:my-subnatSEtable}
\end{table}

From Table \ref{tab:my-subnatSEtable}, the calculated subnational-level standard errors range from 0 to 22 percentage points. The median standard error size across all method is 3.8 percentage points approximately. On average, they tend to be largest for  OC pills where the mean standard error size is approximately 5 percentage points and smallest in implants where the mean standard error size is approximately 4 percentage points. The calculated standard errors of the subnational IPUMS-DHS data are larger than those calculated using the DHS national-level data (Table \ref{tab:my-natSEtable}) . At the national level, the median standard error (across all methods) is 2.23\%. This is almost half the size of the subnational median. Similarly, at the national-level IUDs have the largest mean standard error (4.10\%). This is almost 1\% smaller than the 5.3\% observed for OC pills at the subnational level.

\section{Out-of-sample validation results}

\subsection{Errors and coverage}
We calculate sector specific error terms, $e_{j,s}$, to describe the difference between the observed data point j in sector s, $y_{j,s}$, and the median estimate from the posterior predictive distribution, $y_{j,s}$ such that

\begin{equation}
    e_{j,s} = y_{j,s} - \hat{y}_{j,s}.
\end{equation}

We evaluated the results of the validation using different measures of accuracy and prediction interval calibration.  To evaluate the accuracy of our model, we considered the root mean square error (RMSE) for each sector’s set of estimates. 

\begin{equation}
    \text{RMSE}_{s}= \sqrt{\frac{\sum_{j=1}^{N_{s}} e_{j,s}^2}{N_{s}}},
\end{equation}
where, $N_{s}$ is the number of observations in the sector s. The RMSE can be interpreted as the average error observed across all countries, time points and methods in the test set.  \newline

We also evaluated the mean error (eq. 2.3) and the median absolute errors (eq 2.4). The mean error is the average difference between the observed proportion and true proportion estimated by the model and is an effective measurement of bias within the model. When the mean error is positive, this indicates systematic under-prediction by the model and conversely, a negative mean error indicates that the model is over-estimating the observed data. Median absolute error is the 50th percentile of absolute differences between the observed proportion and true proportion estimated by the model.  Median absolute error captures the overall variation within the model estimates.

\begin{equation}
   \text{Mean error}_{s}= \frac{\sum_{j=1}^{N_{s}} e_{j,s}}{N_{s}}
\end{equation}

\begin{equation}
   \text{Median absolute error}_{s}= \text{Median}(| \boldsymbol{e_{s}}|)
\end{equation}
Where,  $\boldsymbol{e_{s}}$ is is a vector of length $N_{s}$ containing the complete set of errors estimates for all observations belonging to sector s. \newline
 \newline
Coverage assumes that if our model is correctly calibrated, then for each sector the model should be able to capture the test set of out-of-sample observations with 95\% accuracy, where the remaining 5\% of incorrectly estimated observations are approximately evenly distributed above and below the estimated 95\% prediction interval. To examine the bias of our models estimates, we examined the location of the incorrectly estimated test set observations. We consider the proportion of test observations located above and below the estimated prediction intervals. By examining the breakdown of locations, we are evaluating the tendency of the model to under- or over-estimate the test set. If a higher proportion of observations are located below the prediction intervals, this indicates that the model is tending to over-estimate the test set. Similarly, if a higher proportion of the incorrectly estimated observations are located above the prediction intervals, the model is tending to under-estimate the test set.

\subsection{Multi-country national model with cross-method correlations}
The validation results (out-of-sample validation results and a comparison of the model-based estimates to the direct estimates) with a discussion of these results for the multi-country national model with cross-method correlations can be found in the supplementary materials associated with Comiskey et al., (2023).

\subsection{Multi-country subnational model with cross-method correlations}

\begin{table}[H]
\resizebox{\textwidth}{!}{%
\begin{tabular}{|c|c|c|cc|c|c|c|}
\hline
\textbf{Sector} &
  \textbf{\begin{tabular}[c]{@{}c@{}}95\%   \\ coverage\\ \\ (\%)\end{tabular}} &
  \textbf{\begin{tabular}[c]{@{}c@{}}Root mean \\ square error \\ (RMSE)   \\ \\ (\%)\end{tabular}} &
  \multicolumn{2}{c|}{\textbf{\begin{tabular}[c]{@{}c@{}}Proportion of \\ incorrectly estimated \\ observations located \\ above and below \\ the prediction interval \\ (PI) boundary   \\ \\ (\%)\end{tabular}}} &
  \textbf{\begin{tabular}[c]{@{}c@{}}95\% \\ PI width\\ (\%)\end{tabular}} &
  \textbf{\begin{tabular}[c]{@{}c@{}}Mean  \\ error   \\ (\%)\end{tabular}} &
  \textbf{\begin{tabular}[c]{@{}c@{}}Median absolute  \\  error   \\ (\%)\end{tabular}} \\ \hline
\multirow{2}{*}{\textbf{\begin{tabular}[c]{@{}c@{}}Commercial\\ medical\end{tabular}}} &
  \multirow{2}{*}{95.3} &
  \multirow{2}{*}{15.5} &
  \multicolumn{1}{c|}{\textbf{Above}}&
  2.84 &
  \multirow{2}{*}{68.4} &
  \multirow{2}{*}{-2.32} &
  \multirow{2}{*}{7.25} \\ \cline{4-5}
 &
   &
   &
  \multicolumn{1}{c|}{\textbf{Below}} &
  1.90 &
   &
   &
   \\ \hline
\multirow{2}{*}{\textbf{Other}} &
  \multirow{2}{*}{98.1} &
  \multirow{2}{*}{6.13} &
  \multicolumn{1}{c|}{\textbf{Above}} &
  1.18 &
  \multirow{2}{*}{33.3} &
  \multirow{2}{*}{-0.17} &
  \multirow{2}{*}{1.0} \\ \cline{4-5}
 &
   &
   &
  \multicolumn{1}{c|}{\textbf{Below}} &
  0.71 &
   &
   &
   \\ \hline
\multirow{2}{*}{\textbf{Public}} &
  \multirow{2}{*}{97.2} &
  \multirow{2}{*}{15.4} &
  \multicolumn{1}{c|}{\textbf{Above}} &
  0.71 &
  \multirow{2}{*}{72.2} &
  \multirow{2}{*}{2.49} &
  \multirow{2}{*}{7.08} \\ \cline{4-5}
 &
   &
   &
  \multicolumn{1}{c|}{\textbf{Below}} &
  2.13 &
   &
   &
   \\ \hline
\end{tabular}%
}
\caption{Out-of-sample validation results for the test set using multi-country subnational model with cross-method correlations. Coverage is the proportion of the test set observations that are captured within the 95\% prediction interval (PI) produced by the model. We also consider the location of the model-based estimates of the incorrectly estimated test set observations with respect the prediction intervals. Above captures the proportion of test set observations that are higher than their corresponding model-based 95\% prediction interval boundary. Below captures the observations located lower than their corresponding model-based 95\% prediction interval boundary. The RMSE, mean error and the median absolute error are as described. The 95\% PI width reflects the median PI width for each observation estimated by the model.}
\label{tab:tab_6_3}
\end{table}

The multi-country subnational model has been evaluated using various out-of-sample model validation measures to gauge its effectiveness at estimating the method supply shares at a subnational level, while also considering the prediction intervals it uses to produce these estimates. It is performing reasonably well considering the complex nature of the data. It has an overall coverage of approximately 97\%. 
The results for the out-of-sample validation are found in Table \ref{tab:tab_6_3}. The target coverage is 95\%. The model is reasonably well calibrated to the data with the public sector having 97\% coverage and the commercial medical sector having 95\% coverage. The private other sector has 98\% coverage of the test set. The public and private other sectors have coverage of that test set that is slightly higher than expected. The private commercial medical is showing  optimal coverage at 95\%. The root mean square error (RMSE) for the private commercial medical sector and the public sector are both at approximately 15 percentage points. The private other sector has an RMSE of approximately 6 percentage points. We also considered where the incorrectly estimated test set observations lie with respect to the prediction interval bounds to assess the bias of the model. In theory, if the model is unbiased and well calibrated then we would expect an equal proportion of incorrectly estimated observations above and below the prediction interval boundaries. Both the commercial medical and other sector has a higher proportion of observations above the prediction interval boundary. This would imply that the model tends to under-estimate the observations in these sectors. In the public sector, there is a higher proportion of incorrectly estimated test set observations below the prediction interval. This implies that for this sector, the model tends to over-estimate the public sector. The median width of the 95\% prediction intervals is largest in the public sector at 72 percentage points. The private other sector has the smallest median 95\% prediction interval width at 33 percentage points.
The mean error for the private commercial medical is approximately -2 percentage points. The mean error for the public sector is the absolute largest of all three sectors at approximately 2.5 percentage points. The private other sector has a mean error of less than 1 percentage point. The median absolute error of the private commercial medical sector is the largest at approximately 7 percentage points while the median absolute error of the private other sector was the smallest at approximately 1 percentage point. 

\subsection{Comparison of the model-based estimates to the direct estimates}

\begin{figure}[H]
    \centering
    \includegraphics[width=14.5cm]{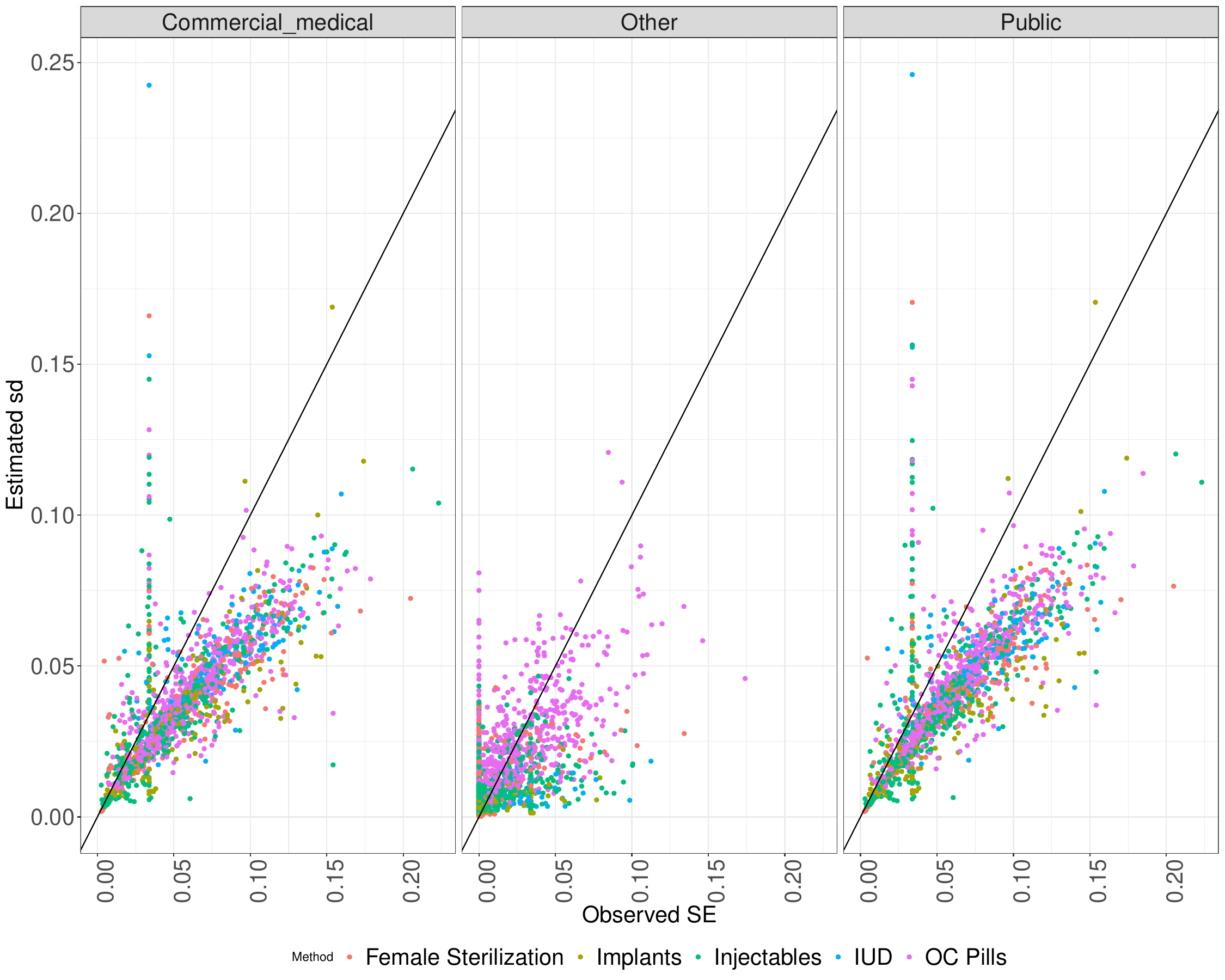}
    \caption{A scatter plot across the three sectors comparing the standard error of the direct estimates to the model-based estimates.}
    \label{fig:subnat_obsSEestSD_plot}
\end{figure}

In Figure \ref{fig:subnat_obsSEestSD_plot}, we consider the observed standard errors calculated using the DHS microdata versus the corresponding standard deviations of the model-based estimates for the proportions. From this figure, we can see that in the commercial medical and public sectors, the estimated standard deviation terms are smaller than the observed standard error terms calculated using the DHS microdata. For both sectors, we can see that most of the observations have standard errors up to 15 percentage points approximately. These same observations when estimated within the model have corresponding standard deviations of approximately up to 10 percentage points. The use of this model results in a considerable reduction in the uncertainty of these observations. The observed SEs of approximately 3 percentage points correspond to the observations where the standard errors were imputed. The outlier observation at 25 percentage points  in both the commercial medical and public sector corresponds to IUDs in Maputo City, Mozambique. IUDs in Maputo only has one observation in 1997, whereas the other methods have more recent survey observations to inform model estimates. The lack of data for IUDs in this instance causes the model estimates to have larger uncertainty. 

\begin{figure}[H]
    \centering
    \includegraphics[width=14.5cm]{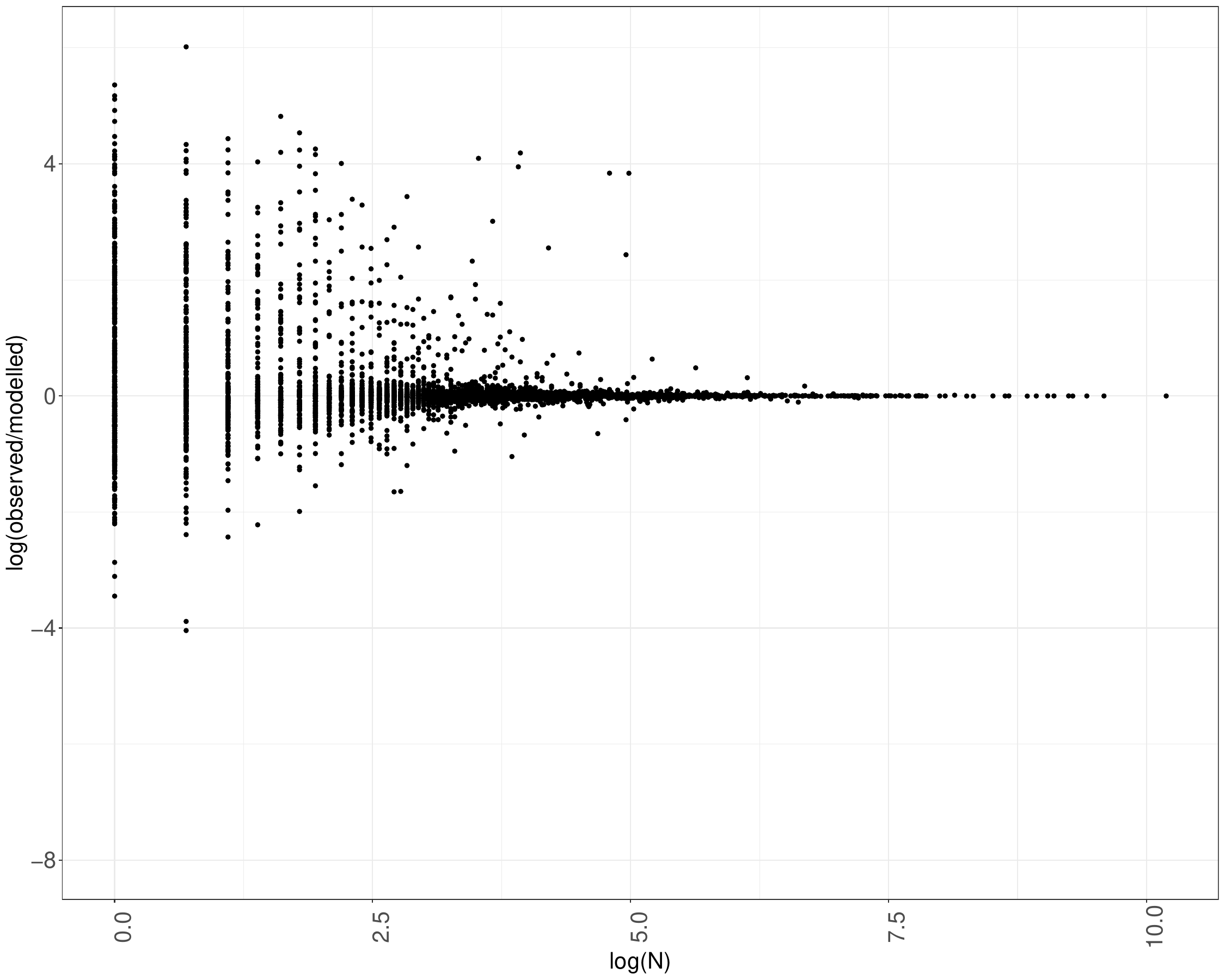}
    \caption{A scatterplot of the estimates with log of the sample size on the x-axis and the log ratio of direct to model-based estimates on the y-axis.}
    \label{fig:subnat_ratioN_plot}
\end{figure}

In Figure \ref{fig:subnat_ratioN_plot}, we consider the sample size with respect to the ratio of observed to estimated proportions, both of which are on the log scale for clarity. From this figure we see that as the sample size increases, the ratio of observed to estimated data point tends towards 1 after approximately log(5), which corresponds approximately to a sample size of 148. Therefore, the model’s ability to capture the observed data point increases as the sample size associated with each observation increases. This aligns with the same property that is seen in many small area estimation models.

\subsection{Model comparison}

\subsubsection{Multi-country subnational model with 0-covariance}

For the subnational multi-country model, we used a 0-covariance model to validate the use of the cross-method correlations within the estimation of $\delta_{q,1:M,s,h}$, the first-order differences between spline coefficients. This approach is similar to that described in Comiskey et al., (2023). In  this instance, the off-diagonal elements of $\Sigma_{\delta_{s}}$ are set to 0. The variance-covariance matrix $\Sigma_{\delta_{s}}$ informs the multi-variate normal prior of $\delta_{q,1:M,s,h}$. 

As before, we describe the first-order differences between spline coefficients, $\delta_{q,1:M,s,h}$, using a Multi-variate Normal prior centred on 0 with variance-covariance matrix $\Sigma_{\delta_{s}}$.

\begin{equation}
    \delta_{q,1:M,s,h} \mid \Sigma_{\delta_{s}}  \sim MVN(\boldsymbol{0}, \Sigma_{\delta_{s}}),
\end{equation} 

such that, $\Sigma_{\delta_{s}}$ is a diagonal matrix with 0 on the off-diagonal elements;
\begin{equation}
    \Sigma_{\delta_s}=\left[\begin{array}{ccccc}
    \sigma_{\delta_{1, s}}^2 & 0 & \ldots & \ldots & 0 \\
    0 & \sigma_{\delta_{2, s}}^2 & \ldots & \ldots & 0 \\
    0 & \ldots & \ldots & \ldots & 0 \\
    \ldots & \ldots & \ldots & \ldots & \ldots \\
    0 & \ldots & \ldots & \ldots & \sigma_{\delta_{M, s}}^2
\end{array}\right].
\end{equation}

Having 0-covariance between the variance terms of $\Sigma_{\delta_{s}}$ implies that the rates of change in method supplies act independently of one another. We compare the validation results for this simpler 0-covariance model with the model using cross-method correlations to investigate the impact of including cross-method correlations in the model estimation process.

\begin{table}[H]
\resizebox{\textwidth}{!}{%
\begin{tabular}{|c|c|c|cc|c|c|c|}
\hline
\textbf{Sector} &
  \textbf{\begin{tabular}[c]{@{}c@{}}95\%   \\ coverage\\ \\ (\%)\end{tabular}} &
  \textbf{\begin{tabular}[c]{@{}c@{}}Root mean \\ square error \\ (RMSE)   \\ \\ (\%)\end{tabular}} &
  \multicolumn{2}{c|}{\textbf{\begin{tabular}[c]{@{}c@{}}Proportion of \\ incorrectly estimated \\ observations located \\ above and below \\ the prediction interval \\ (PI) boundary   \\ \\ (\%)\end{tabular}}} &
  \textbf{\begin{tabular}[c]{@{}c@{}}95\% \\ PI width\\ (\%)\end{tabular}} &
  \textbf{\begin{tabular}[c]{@{}c@{}}Mean \\ error   \\ (\%)\end{tabular}} &
  \textbf{\begin{tabular}[c]{@{}c@{}}Median absolute \\ error   \\ (\%)\end{tabular}} \\ \hline
\multirow{2}{*}{\textbf{\begin{tabular}[c]{@{}c@{}}Commercial\\ medical\end{tabular}}} &
  \multirow{2}{*}{96.4} &
  \multirow{2}{*}{17.4} &
  \multicolumn{1}{c|}{\textbf{Above}} &
  2.37 &
  \multirow{2}{*}{68.9} &
  \multirow{2}{*}{-3.02} &
  \multirow{2}{*}{7.35} \\ \cline{4-5}
 &
   &
   &
  \multicolumn{1}{c|}{\textbf{Below}} &
  1.18 &
   &
   &
   \\ \hline
\multirow{2}{*}{\textbf{Other}} &
  \multirow{2}{*}{98.1} &
  \multirow{2}{*}{6.77} &
  \multicolumn{1}{c|}{\textbf{Above}} &
  1.42 &
  \multirow{2}{*}{33.4} &
  \multirow{2}{*}{0.768} &
  \multirow{2}{*}{0.976} \\ \cline{4-5}
 &
   &
   &
  \multicolumn{1}{c|}{\textbf{Below}} &
  0.474 &
   &
   &
   \\ \hline
\multirow{2}{*}{\textbf{Public}} &
  \multirow{2}{*}{97.9} &
  \multirow{2}{*}{14.8} &
  \multicolumn{1}{c|}{\textbf{Above}} &
  0.474 &
  \multirow{2}{*}{72.8} &
  \multirow{2}{*}{2.30} &
  \multirow{2}{*}{7.21} \\ \cline{4-5}
 &
   &
   &
  \multicolumn{1}{c|}{\textbf{Below}} &
  1.66 &
   &
   &
   \\ \hline
\end{tabular}%
}
\caption{Out-of-sample validation results for the test set using multi-country subnational model with 0-covariance. Coverage is the proportion of the test set observations that are captured within the 95\% prediction interval (PI) produced by the model. We also consider the location of the model-based estimates of the incorrectly estimated test set observations with respect the prediction intervals. Above captures the proportion of test set observations that are higher than their corresponding model-based 95\% prediction interval boundary. Below captures the observations located lower than their corresponding model-based 95\% prediction interval boundary. THE RMSE, mean error and the median absolute error are as described. The 95\% PI width reflects the median PI width for each observation estimated by the model.}
\label{tab:tab_6_4}
\end{table}

The coverage of the 0-covariance model (Table \ref{tab:tab_6_4}) is higher than that of the cross-method correlation model (Table \ref{tab:tab_6_3}). The 0-covariance model has 98\% coverage in both the public and private other sectors. The commercial medical sector has 96\% coverage (Table \ref{tab:tab_6_4}). The coverage of the model with cross-method correlations is Commercial medical = 95\%, Other = 98\%, Public = 97\% (Table \ref{tab:tab_6_3}).

To evaluate the bias and variance produced by the 0-covariance and cross-method correlation models, we consider the mean errors, median absolute errors (MAE) and root mean square errors (RMSE). Across all three sectors, the RMSE of the 0-covariance model is larger than the cross-method correlation model. The private commercial medical sector has the largest RMSE with an average error of approximately 17 percentage points(Table \ref{tab:tab_6_4}). The RMSE of the private commercial medical sector in the cross-method correlation model is approximately 2 percentage points smaller at 15 percentage points (Table \ref{tab:tab_6_3}). In both models, the private other sector has the smallest RMSE. In the 0-covariance model, it is approximately 7 percentage points whereas in the cross-method correlation model it is approximately 6 percentage points. Overall, the cross-method correlation model performs better in this model validation measure than the 0-covariance model. In both models the mean error on the private commercial medical sector is negative (-3.02 percentage points in the 0-covariance model and -2.32 percentage points in the cross-method correlation model) and the mean error of the private other and public sectors are positive. This implies that both models over-predict the test set of the commercial medical sector and under-predict the remaining two sectors. The median absolute errors (MAE) of both models are very similar. Both the 0-covariance and cross-method correlation models see the largest MAE in the private commercial medical sector (7.35 percentage points in the 0-covariance model and 7.25 percentage points in the cross-method correlation model). In both models, the private other sector has an MAE of less than 1 percentage point (0.1 percentage points in the 0-covariance model and 0.1 percentage points in the cross-method correlation model).

When considering the median prediction interval widths, we see that the cross-method correlation model has slightly smaller sized prediction interval widths as compared to the 0-covariance model for the commercial medical and public sectors (68 percentage points and 72 percentage points Table \ref{tab:tab_6_3}; 69 percentage points and 73 percentage points Table \ref{tab:tab_6_4}). The median prediction interval width of the private other sector is the same in both models at 33 percentage points.

Lastly, when considering the location of the incorrectly estimated test set observations, the cross-method correlation model and 0-covariance models both tend to over-estimate the public sector (as there is higher proportion of incorrectly estimated observations below the prediction interval) and under-estimate the private commercial medical and private other sectors (as there is higher proportion of incorrectly estimated observations above the prediction interval) (Table \ref{tab:tab_6_3}) (Table \ref{tab:tab_6_4}). 

Overall, the multi-country subnational model with cross-method correlations is the most suitable model to describe this complex data. It captures the complex shape and relationships without over-fitting it or missing the shape. It incorporates information regarding the correlations between the rates of change across the contraceptive methods. The coverage, RMSE and median 95\% prediction intervals widths produced by the cross-method correlation model are similar but slightly better than those of the 0-covariance model. The strength of the full model is seen in the absence of data for a particular contraceptive method, where model estimates can still be informed by the behaviour of related methods to produce realistic estimates.

\section{Single-country national and subnational model validation}
We validate the single country models indirectly using the multi-country model estimates. The idea here is that, by comparing the median estimates of the single-country model to the validated multi-country model estimates when can get a gauge of the reliability of our single-country model estimates. If the single-country model estimates align with the multi-country model estimates, then they too are validated.

\section{Single-country national model validation}

\begin{figure}[H]
    \centering
    \includegraphics[width=14.5cm]{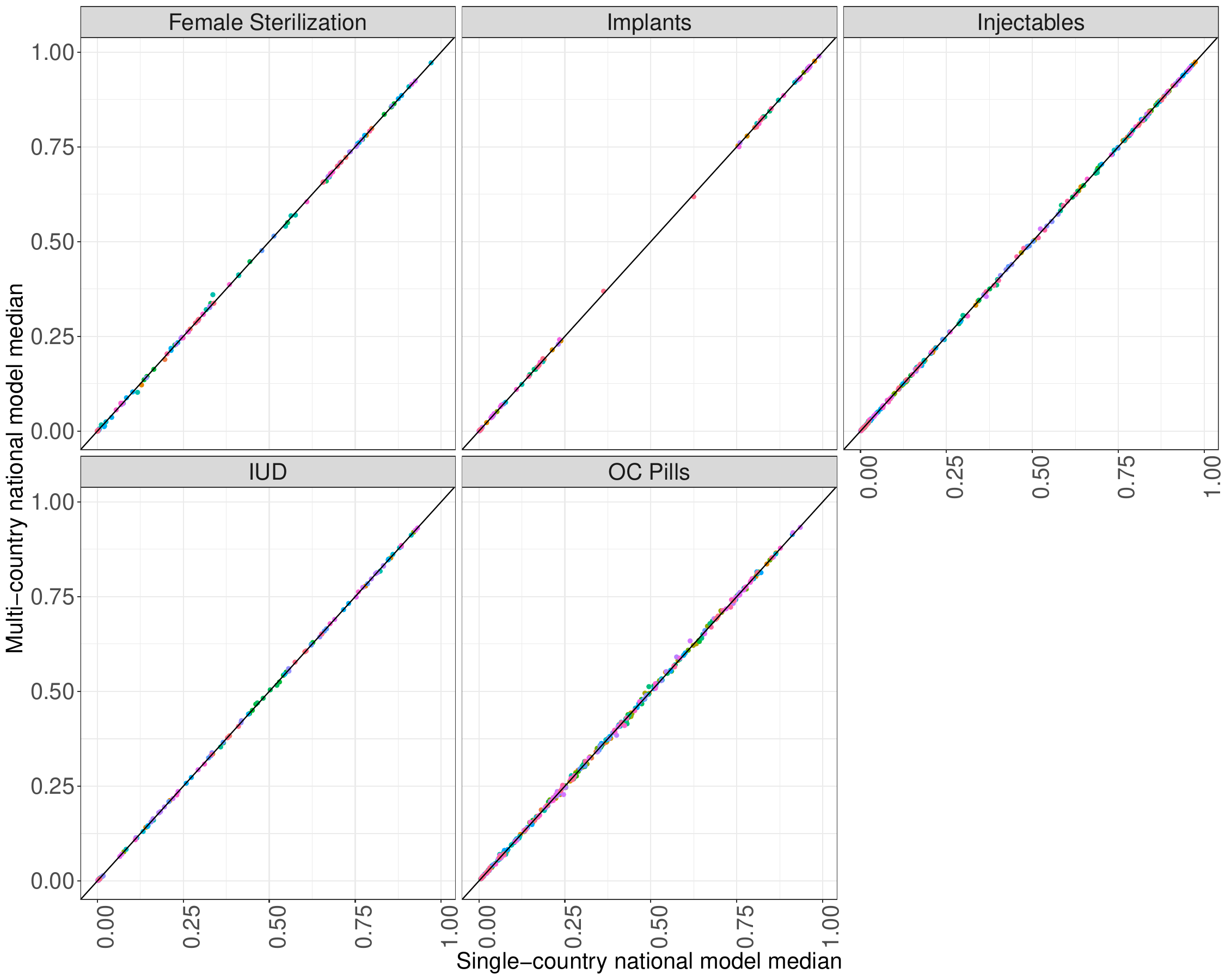}
    \caption{A scatterplot comparing the median estimates of each country, method, sector, and time point estimated by the single-country model (x-axis) and the multi-country model (y-axis). Each panel represents a different method and each colour represents a country. The diagonal line capture the 1:1 agreement between the two modelling approaches. }
    \label{fig:nat_mc_sc_median_validation}
\end{figure}

In Figure \ref{fig:nat_mc_sc_median_validation}, it is clear that the single-country national model median estimates align approximately with the multi-country national median estimates. Therefore, we can conclude that the single-country national model estimates are as valid and reliable as those estimated by the multi-country national model.

\section{Single-country subnational model validation}

\begin{figure}[H]
    \centering
    \includegraphics[width=14.5cm]{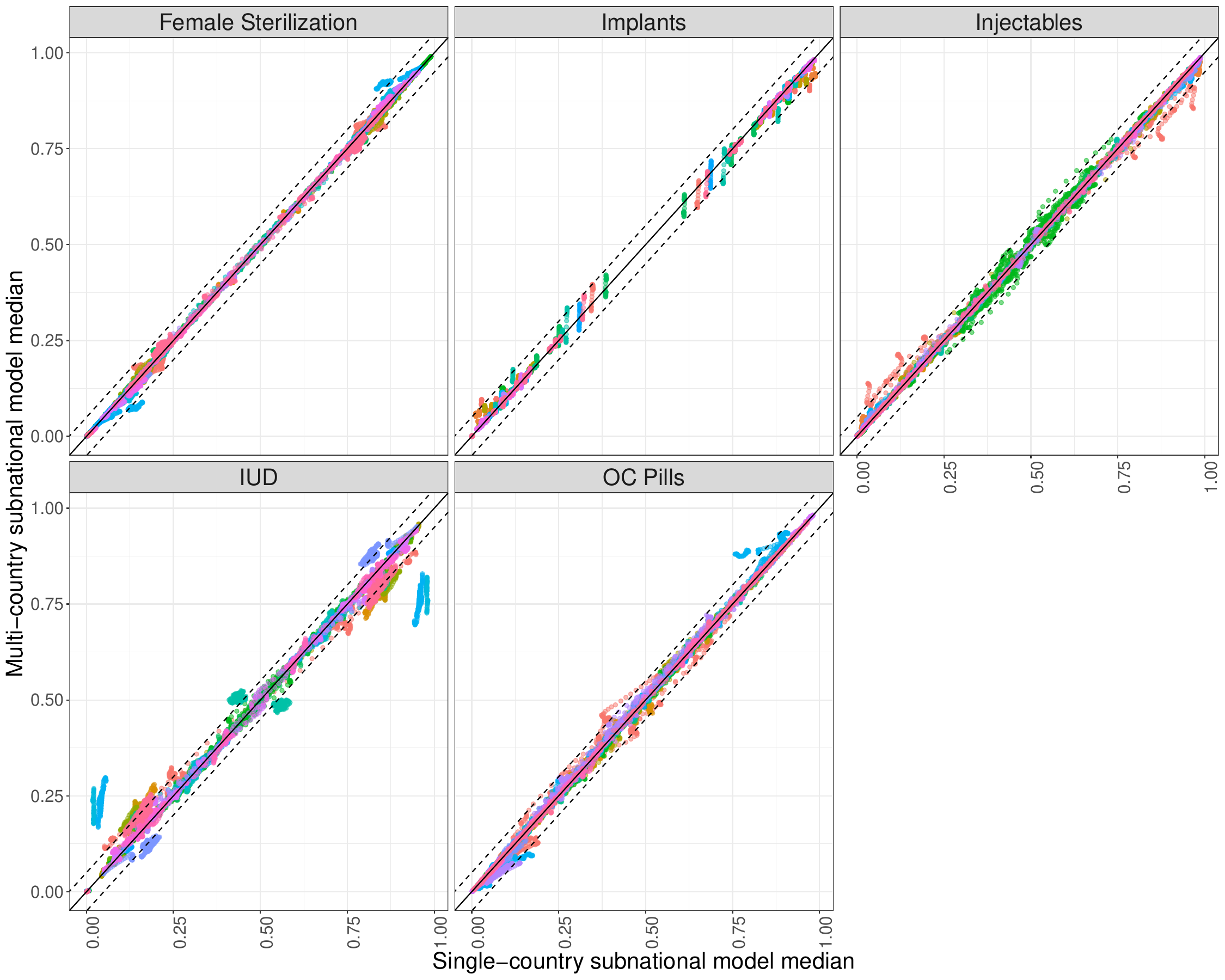}
    \caption{A scatterplot comparing the median estimates of each country, method, sector, and time point estimated by the single-country subnational model (x-axis) and the multi-country subnational model (y-axis). Each panel represents a different method and each colour represents a country. The outer dashed lines represent the +5\% and -5\% from complete agreement between the two models. The diagonal solid line capture the 1:1 agreement between the two modelling approaches. }
    \label{fig:subnat_mc_sc_median_validation}
\end{figure}

In Figure \ref{fig:subnat_mc_sc_median_validation}, we can see that when comparing the single-country estimates to the multi-country estimates, the majority of observations fall inside the $\pm5\%$ boundary. This means that the estimates and projections produced by both models have a difference of up to $\pm5\%$. There are few observations in IUDs that are outliers to this. These belong to Mozambique, where there is only one survey in 1997 taken in the City of Maputo. Therefore, we can conclude that the single-country subnational model median estimates are as valid and reliable as those estimated by the multi-country subnational model.

\end{document}